\documentclass[aps,showpacs,preprint]{revtex4}
\usepackage{epsfig,amssymb,bm,graphics}

\newcommand*{\fs}[1]{#1\!\!\!/}

\begin{document}


\title{
Exclusive charm production in $\bar pp$ collisions at 
$\boldmath  \sqrt{s} \lesssim 15 {\rm GeV}$}

 \author{A.I.~Titov$^{a,b}$ and B.~K\"ampfer$^{a,c}$}
 \affiliation{
 $^a$Forschungzentrum Dresden-Rossendorf, 01314 Dresden, Germany\\
 $^b$Bogoliubov Laboratory of Theoretical Physics, JINR,
  Dubna 141980, Russia\\
 $^c$ Institut f\"ur Theoretische Physik, TU~Dresden, 01062 Dresden,
 Germany
 }


\begin{abstract}
 We discuss the open charm production in peripheral reactions
 $\bar pp\to \bar Y_cY_c$ and $\bar pp\to M_c\bar M_c$, where
 $Y_c$ and $M_c$ stand for $\Lambda_c^+,\Sigma_c^+$ and
 $D,D^*$, respectively, at $\sqrt{s}\lesssim 15$~GeV,
 which corresponds to the energy range of FAIR.
 Our consideration is based on the topological
 decomposition of the planar quark and diquark diagrams which allows
 to estimate consistently meson and baryon
 exchange trajectories and energy scale parameters as well.
 The spin dependance is determined by the
 effective interaction of lowest exchanged resonance.
 Unknown parameters are fixed by an independent analysis of open
 strangeness production in
 $\bar pp\to \bar YY$ and $\bar pp\to \bar KK$
 reactions and of SU(4) symmetry.
 We present the corresponding cross sections and longitudinal double-spin asymmetries
 for exclusive binary reactions with open charm mesons and baryons
 in the final state.
 The polarization observables have a non-trivial $t$ and $s$ dependence
 which is sensitive to details of the open charm production mechanism.
\end{abstract}

\pacs{12.38Bx, 13.85Ni, 13.88.+e}

\maketitle


\section{Introduction}

Open charm production will be one of the major topics of the
hadron and heavy-ion programme at FAIR~\cite{FAIR}.
On the one hand, charm spectroscopy will be addressed by
the PANDA collaboration \cite{PANDA}, while the CBM collaboration
\cite{CBM} will exploit
charmed particles as probes of the nuclear medium at maximum compression.
For both large-scale experiments at FAIR one needs to know the
properties of charmed baryons and mesons as well as their
production processes in elementary $pp$ and $\bar pp$ reactions.
For this the opportunities at FAIR are promising, as for instance,
the PAX collaboration~\cite{PAX}) envisages the use a polarized
anti-proton beam. This offers the chance to study in depth the
mechanism of open charm production at the moderate energies from
threshold to $\sqrt{s} \lesssim 15 {\rm GeV}$. In this energy
range the phenomenology of charm production is not well
established. In present paper we select one important problem of
this wide field, namely, the analysis of exclusive binary
reactions $\bar pp\to \bar Y_cY_c$, $\bar pp\to D\bar D$, $\bar
pp\to D\bar D^*$ etc., in peripheral collisions in the mentioned
energy range.

Since the initial energy  is not asymptotically high, the widely
used models for the heavy quark production based on perturbative QCD (see for
example Ref.~\cite{Nason1989,Vogt,Kneesch}) are not applicable,
and an essential improvement by including high order corrections
is needed~\cite{Florian,Riedl}. Another severe problem is related
to the dynamics of charm productions. In the popular QCD models,
the $c$ quark is produced through gluon fragmentation. For $c$
production in peripheral collisions such a gluon must have a large
momentum (large $x \sim 1$), i.e., much larger than its average
value in a nucleon $x\lesssim 0.2$, and therefore, this mechanism
is strongly suppressed.

It is expected that the mechanisms of peripheral charm production
in $\bar pp\to D\bar D$ and $\bar pp\to \bar Y_cY_c$ reactions are
similar to the strangeness production in $\bar pp\to \bar KK$ and
$\bar pp\to \bar YY$  reactions, respectively, which were
described successfully in terms of Regge pole
models~\cite{Storrow_PR,Saleem,Plaut,Sadoulet} with certain baryon
and vector meson exchange trajectories. However, a direct
extrapolation of such  models to the charm production is faced by
a problem. First, the linear Regge trajectories leads to large
negative values of the corresponding intercepts
$\alpha_{\Lambda_c}(0)\sim-4.5$ and $\alpha_{D^*}(0)\sim-2$. This
would result in a suppression of the charm production in  peripheral
reactions which contradicts to the corresponding data on inclusive
charm production. This means, in turn, that the trajectories connected
to masses and spins of the charmed hadrons must be essentially
non-linear (cf.\
Refs.~\cite{Brisudova2000,BurakovskyGoldmanHorvitz}. Another
problem is the estimate of the energy scale parameter
in the Regge pole propagator, which also affects much the cross
sections.

\begin{figure}[ht]
 \includegraphics[width=0.4\columnwidth]{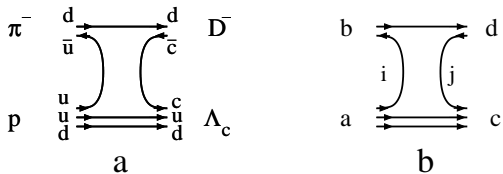}
\caption{\small{Diagrammatic  representation of
planar diagram for reaction  $\pi^-p\to D^-\Lambda_c$ (a)
and for more general case (b).}}
\label{Fig:1}
\end{figure}

Therefore, it seems to be interesting and important to use a model
based essentially on a non-perturbative QCD background being
reliable for describing the peripheral reactions. Such
an approach was developed 
in Refs.~\cite{KaidalovZP,BoreskovKaidalov,KaidalovPiskunova} and
applied for the evaluation of cross sections of the exclusive
$\Lambda_c$ production in $\pi p$ and $pp$ collisions. The binary
$\pi^-p\to D^-\Lambda_c$ exclusive process plays an important role
in this consideration~\cite{BoreskovKaidalov}. The model for this
reaction is based on quark-gluon string dynamics, assuming the
annihilation of a $q\bar q$ pair in the interaction, the formation of a
$q\bar q$ color tube with subsequent decay to the observed
hadrons (see also Refs.~\cite{Low,Nussinov}). Schematically, the
process $\pi^-p\to D^-\Lambda_c$ is described by the planar
diagram shown in Fig.~\ref{Fig:1}~(a).
A more general case is exhibited in Fig.~\ref{Fig:1}~(b).
The assumption of the formation and decay of color-gluon strings
allows to construct the space-time evolution of the process and
to obtain the factorization condition, where the imaginary part
of the amplitude of the
process $a b\to c d$ is expressed via a product of the
probabilities $w_{ab}$ and $w_{cd}$ of the elastic scattering of
$ab\to ab$ and $cd\to cd$, respectively. This gives a consistent
prescription for evaluating the parameters of the amplitude
(trajectories and energy scale parameters) of the non-diagonal
transition $ab \to cd$.

The aim of the present paper is to extend the results of
Refs.~\cite{KaidalovZP,BoreskovKaidalov,KaidalovPiskunova} for
exclusive charm production in the $\bar pp$ collisions. In our
consideration we analyze simultaneously the open charm and
open strangeness production. We are going to consider the reactions
$\bar pp\to \bar\Lambda\Lambda$ and $\bar pp\to
\bar\Lambda_c\Lambda_c$, $\bar pp\to \bar KK$ and  $\bar pp\to
D\bar D$ etc. The strangeness production has its own interest, but
on the other hand, some available (although relatively old)
experimental data allow to fix the unknown parameters of the model
and get absolute values of the cross sections of the open charm
production. We also analyze the double longitudinal asymmetry
which will be accessible in mentioned the FAIR PAX experiment due
to the polarized anti-proton beam. The spin dependence of the
amplitudes is generated by the symmetry of the $PNY$ and $VNY$
interactions ($P=K,D$ and $V=K^*,D^*...$) which was widely used in
various studies (see e.g.\ Refs.~\cite{TKR,Guidal}).

For completeness, we mention that some aspects of the
inclusive charm production in terms of quark-gluon string model
were discussed in Refs.~\cite{Lykasov,Arakelian}; polarization
effects in open charm photo-production were considered in
Ref.~\cite{TG_R}; propagation of charmed hadrons in the
nuclear medium were analyzed in
Refs.~\cite{Cassing1,Cassing2,Sibirtsev}; open charm
production in relativistic nucleus-nucleus collisions at wide
energy region was analyzed in Ref.~\cite{Cassing3}.

 Our paper is organized as follows.
 In Sec.~II we analyze
 the strangeness production in the reactions
 $\bar pp\to \bar\Lambda\Lambda$, $\bar pp\to \bar\Lambda\Sigma^0$,
 and $\bar pp\to \bar\Sigma^0\Sigma^0$, and the open charm production in
 $\bar pp\to \bar\Lambda^+_c\Lambda^+_c$,
 $\bar pp\to \bar\Lambda^+_c\Sigma^+_c$, and
 $\bar pp\to \bar\Sigma^+_c\Sigma^+_c$ processes,
 where the dominant contribution comes from the $K^*$ and $D^*$ exchange
 trajectories, respectively. First, we note equations for
 the invariant amplitudes and then discuss our results for the
 differential cross sections and the longitudinal asymmetry.
 In Sec.~III we provide a similar analysis for
 the $\bar pp\to \bar KK$ and $\bar pp\to D\bar D$ reactions,
 assuming the dominance of strange
 and charmed baryon exchange trajectories.
 The reactions  $\bar pp\to \bar KK^*$ and
 $\bar pp\to D\bar D^*$ are discussed in Sec.~IV.
 The summary is given in Sec.~V.

\section{Reactions
$\boldmath \bar pp\to \bar YY$ and $\boldmath \bar pp\to\bar Y_cY'_c$}

In this section, we discuss strange and charmed baryon-antibaryon
production in peripheral $\bar pp$ collisions. For the sake of simplicity, we
consider the exclusive production of $\bar\Lambda\Lambda$ and
$\bar\Lambda_c\Lambda_c$ pairs. The generalization for reactions with
$\bar\Lambda\Sigma$, $\bar\Sigma\Lambda$, $\bar\Sigma\Sigma$ final
states may be done in a straightforward manner.

The corresponding planar diagrams for $\bar\Lambda\Lambda$ and
$\bar\Lambda_c\Lambda_c$ are depicted in Fig.~2~(a) and (b).
\begin{figure}[ht]
   \includegraphics[width=0.4\columnwidth]{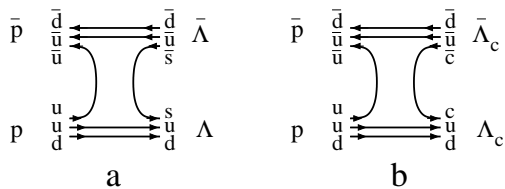}
\caption{\small{
Planar diagram for the reactions $\bar p\to \Lambda\Lambda$ (a)
and $\bar p\to \Lambda_c\Lambda_c$ (b).}}
\label{Fig:2}
\end{figure}

\subsection{Reaction
$\boldmath \bar pp\to \bar\Lambda\Lambda$}

Following Ref.~\cite{BoreskovKaidalov} we assume that the
amplitude of the reaction $\bar pp\to \bar\Lambda\Lambda$ has the
form of a Regge pole amplitude, dominated by the $K^*$ exchange
trajectory,
\begin{eqnarray}
 T^{\bar pp\to\bar \Lambda\Lambda}_{m_f n_f;m_i,n_i}
 =C(t){\cal M}^{\bar pp\to\bar \Lambda\Lambda}_{m_f n_f;m_i,n_i}(s,t)
\frac{g_{K^*N\Lambda}^2}{s_0}\,
\Gamma(1-\alpha_{\bar sq}(t))\,
\left(-\frac{s}{s_{\bar pp:\bar\Lambda\Lambda}}
\right)^{\alpha_{\bar sq}(t)-1},
\label{E1}
\end{eqnarray}
where $m_i,\,m_f$, $n_i$ and $n_f$ are the spin projections of
$p,\,\Lambda$, $\bar p$, and $\bar\Lambda$, respectively, $q$
stands for $u$ and $d$ quarks, $\alpha_{\bar sq}(t)$ is the
$K^{*+}$ trajectory, $g_{K^*N\Lambda}^2$ is the coupling constant
of the $K^*N\Lambda$ interaction, and $s_0=1$~GeV is an universal
scale parameter. The spin dependence is accumulated in the
amplitude ${\cal M}$ which is a smooth function of the Mandelstam variables $s$ and $t$.
In the limit of $s \to \infty$ one has ${\cal M} \propto s $. The explicit form
of ${\cal M}$ will be defined later on. The overall residual function
$C(t)$ will be found from a comparison with available experimental data.

In our consideration we use the nonlinear representation for the
meson trajectories developed in Ref.~\cite{Brisudova2000}
\begin{eqnarray}
\alpha(t)=\alpha(0)+\gamma(\sqrt{T}-\sqrt{T-t}), \label{E4}
\end{eqnarray}
where $\gamma=3.65$~GeV$^{-1}$ is the universal parameter (a
universal slope in the asymptotic region), and $T\gg1$~GeV$^2$ is the scale
parameter, being special for each trajectory. In the diffractive
region with $-t\ll T$, the linear approximation
\begin{eqnarray}
\alpha(t)=\alpha(0)+\alpha't, \label{E5}
 \end{eqnarray}
is valid with $\alpha'=\gamma/2\sqrt{T}$.

The intercept $\alpha_{\bar sq}(0)$ and the slope  $\alpha_{\bar
sq}'$ of the trajectory for the non-diagonal transition are
related to the corresponding parameters for diagonal transitions
as following~\cite{KaidalovZP,Brisudova2000}
\begin{eqnarray}
2\alpha_{\bar sq}(0)&=&\alpha_{\bar qq}(0)+\alpha_{\bar ss}(0)~,
\label{E2}\\
{2}/{\alpha'_{\bar sq}}&=& 1/{ \alpha_{\bar qq}'}+{1}/{
\alpha_{\bar ss}'}~, \label{E3}
 \end{eqnarray}
where $\alpha_{\bar qq}(t)$ and $\alpha_{\bar ss}(t)$ are the
$\rho$ and $\phi$ meson trajectories, respectively.

In our numerical calculations we employ
\begin{eqnarray}
\alpha_{\rho}(0)&=& 0.55,\qquad\sqrt{T_\rho}=2.46~{\rm GeV},
\qquad\alpha_{\rho}'\simeq0.742~{\rm GeV}^{-2}~, \nonumber\\
\alpha_{K^*}(0)&=& 0.414, \qquad \sqrt{T_{K^*}}=2.58~{\rm GeV},
\qquad\alpha_{K^*}'\simeq0.71~{\rm GeV}^{-2}~,\nonumber\\
\alpha_{\phi}(0)&=& 0.28, \qquad \sqrt{T_\phi}\simeq2.70~{\rm GeV},
\qquad\alpha_{\phi}'\simeq0.676~{\rm GeV}^{-2}~,
\label{E6}
\end{eqnarray}
where the $\rho$ and $K^*$ trajectories are taken as input
according to Ref.~\cite{Brisudova2000}.

The energy scale parameter $s_{\bar pp:\bar \Lambda\Lambda}$ in
Eq.~(\ref{E1}) is related to the corresponding scale parameters
for the diagonal transitions $\bar pp\to \bar pp$, ($s_{\bar pp} $) and $\bar
\Lambda\Lambda\to\bar \Lambda\Lambda$, ($s_{\bar \Lambda\Lambda}$) as
 \begin{eqnarray}
 \left(s_{\bar pp:\bar
 \Lambda\Lambda}\right)^{2(\alpha_{K^*}(0)-1)} = \left(s_{\bar
 pp}\right)^{\alpha_{\rho}(0)-1}\times
 \left(s_{\bar\Lambda\Lambda}\right)^{\alpha_{\phi}(0)-1}~.
 \label{E7}
 \end{eqnarray}
The scale parameter for the diagonal transition $s_{ab}$ is determined
by the sum of the transversal masses of the constituent quarks~\cite{BoreskovKaidalov}
 \begin{eqnarray}
  s_{ab}=\left(\sum\limits_{i}^{n_a} {M_{i}}_\perp\right)
  \left(\sum\limits_{j}^{n_b} {M_{j}}_\perp\right)
  \label{E8}
 \end{eqnarray}
 with ${M_{q}}_\perp\simeq0.5$~GeV, ${M_{s}}_\perp\simeq0.6$~GeV,
 and ${M_{c}}_\perp\simeq1.6$~GeV. This leads to the following
 values for the scale factors: $s_{\bar pp}\simeq2.25$~GeV$^2$,
 $s_{\bar \Lambda\Lambda}\simeq2.56$~GeV$^2$,
 and $s_{\bar pp:\bar\Lambda\Lambda}\simeq2.43$~GeV$^2$.

We assume that the spin dependence of the  amplitude  in Eq.~(\ref{E1})
is determined by the symmetry of the $N\Lambda K^*$ interaction given
by the effective Lagrangian in the conventional form
\begin{eqnarray}
 {\cal L}_{K^*NY}
 =
  - \bar{Y}\left(\gamma_\mu
 -\frac{\kappa_{K^*NY}}{M_N+M_Y}
\sigma_{\mu\nu} \right) {N}\partial^\nu K^{*\mu} +{\rm h.c.}~,
\label{E9}
\end{eqnarray}
where $N,Y$ and $K^*$ denote the nucleon, hyperon and the $K^*$
meson fields, respectively, $Y$ stands for $\Lambda$, $\Sigma$,
etc., and $\kappa$ is the tensor coupling strength.

Using this form, one can get the amplitude  ${\cal M}$ in
Eq.~(\ref{E1}) as
\begin{eqnarray}
{\cal M}^{\bar pp\to\bar \Lambda\Lambda}_{m_f n_f;m_i n_f}(s,t)
&=&{\cal N}(s,t)\,\Gamma^{(p)\,\mu}_{m_fm_i}\,\,
\Gamma^{(\bar p)\,\nu}_{n_f n_i}\,\,
(-g_{\mu\nu} + \frac{q_\mu q_\nu}{q^2})~,
\label{E10}
\end{eqnarray}
where $q$ is momentum transfers in the $p\Lambda K^*$ vertex:
$q=p_p-p_\Lambda$, with $p_p$ and $p_\Lambda$ as four-momenta of
the incoming proton and outgoing $\Lambda$. The functions
$\Gamma^{(p,\bar p)}$ read
\begin{eqnarray}
\Gamma^{(p)}_\mu&=& \bar u_{\Lambda}
\left((1+\kappa_{K^*N\Lambda})\gamma_\mu -
\kappa_{K^*N\Lambda}\frac{(p_p+p_\Lambda)_\mu}{M_N+M_\Lambda})
\right)\,u_{p}~,\nonumber\\
\Gamma^{(\bar p)}_\mu&=& \bar v_{\bar p}
\left((1+\kappa_{K^*N\Lambda})\gamma_\mu +
\kappa_{K^*N\Lambda}\frac{(p_{\bar p}
+p_{\bar\Lambda})_\mu}{M_N+M_\Lambda}) \right)\,v_{\bar \Lambda}~.
\label{E11}
\end{eqnarray}
The normalization factor ${\cal N}(s,t)$ eliminates additional $s$
and $t$ dependence provided by the Dirac structure in
Eq.~(\ref{E10}) which is beyond the Regge parametrization:
\begin{eqnarray}
{\cal N}(s,t)&=&
\frac{F_{\infty}(s)}{F(s,t)},\qquad F_{\infty}(s)=2s~, \nonumber\\
F^2(s,t)&=& {\rm Tr}
\left(\Gamma^{(p)\,\mu}{\Gamma^{(p)\,\mu'}}^\dagger\right) {\rm
Tr}\left(\Gamma^{(\bar p)\,\nu} {\Gamma^{(\bar
p)\,\nu'}}^\dagger\right)\, (g_{\mu\nu} - \frac{q_\mu q_\nu}{q^2})
(g_{\mu'\nu'} - \frac{q_{\mu'} q_{\nu'}}{q^2})~. \label{E12}
\end{eqnarray}

For the $NYK^*$ coupling constants we use the average values of
the Nijmegen potential~\cite{Stoks1999}: $g_{K^*NY}=-5.18$,
$\kappa_{K^* NY}=2.79$ for $Y=\Lambda$ and ($-3.29$, $-0.91$) for
$Y=\Sigma^0$.

The cross section is related to the invariant amplitude
of Eq.~(\ref{E1}) as
\begin{eqnarray}
\frac{d\sigma}{dt}
=\frac{1}{16\pi(s-4M_N^2)^2}|T_{fi}|^2~,
\label{E13}
\end{eqnarray}
where summing and averaging over the spin projection in initial
and the final state is provided. We will also discuss the
longitudinal double spin asymmetry, defined as
\begin{eqnarray}
{\cal A}
=\frac{d\sigma^\leftrightarrows - d\sigma^\rightrightarrows}
 {d\sigma^\leftrightarrows  + d\sigma^\rightrightarrows},
\label{E14}
\end{eqnarray}
where the symbols $\leftrightarrows$ and $\rightrightarrows$
correspond to the anti-parallel and parallel spin projections of
incoming $p$ and $\bar p$ with respect to the quantization axis
chosen along the proton momentum in the center-of-mass system (c.m.s.).

The generalization for reactions $\bar pp\to\bar\Lambda\Sigma$, $\bar
pp\to\bar\Sigma\Lambda$, and $\bar pp\to\bar\Sigma\Sigma$ is accomplished
by the substitution of $M_\Lambda\to M_\Sigma$, $g_{K^*N\Lambda}\to
g_{K^*N\Sigma}$ and $\kappa_{K^*N\Lambda}\to \kappa_{K^*N\Sigma}$
in Eqs.~(\ref{E1}) and (\ref{E11}).

\subsection{Reaction
$\boldmath \bar pp\to \bar\Lambda_c\Lambda_c$}

In this case, the amplitude is defined by Eq.~(\ref{E1}) with the
obvious substitution $\Lambda\to \Lambda^+_c\equiv\Lambda_c$,
$\Sigma^0\to\Sigma_c^+\equiv\Sigma_c$, $K^*\to \bar D^*$,
$\alpha_\phi\to \alpha_{J/\psi}$ etc. As a first approximation, we
assume the validity of SU(4) symmetry and, therefore, the coupling
constants of the $D^{*}NY_c$ interaction are chosen to be the same
as for the case of $K^*NY$ interaction. The corresponding
trajectory and the energy  scale parameters read
\begin{eqnarray}
\alpha_{D^*}(0)&=& -1.02,\qquad\sqrt{T_{D^*}}=3.91~{\rm GeV},
\qquad\alpha_{D^*}'\simeq0.467~{\rm GeV}^{-2}~,\nonumber\\
\alpha_{J/\psi}(0)&=&-2.60,\qquad\sqrt{T_{J/\psi}}\simeq5.36~{\rm GeV},
\qquad\alpha_{J/\psi}'\simeq0.34~{\rm GeV}^{-2}~,\nonumber\\
s_{\bar pp:\bar\Lambda_c\Lambda_c}&\simeq&5.98~{\rm GeV}^2.
\label{E15}
 \end{eqnarray}

\subsection{Results}

\subsubsection{Differential cross sections}

\begin{figure}[ht!]
    \parbox{.5\textwidth}{
    \includegraphics[width=0.36\columnwidth]{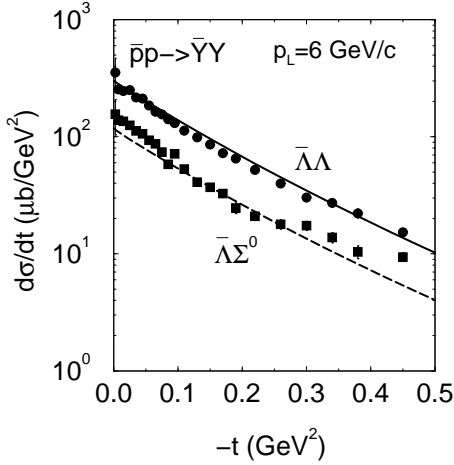}
}
\hfill
\parbox{.4\textwidth}{
   \caption{\small{
   Differential cross section of the reactions $\bar p p\to \bar \Lambda$
   (solid curve) and $\bar pp\to\bar\Lambda\Sigma^0$ (dashed curve)
   as a function of the momentum transfer $t$ at $p_L=6$~GeV.
  The experimental data are taken from  Ref.~\protect\cite{Becker1978}.
\label{Fig:3} }}}
  \end{figure}

 Consider first the strange hyperon production $\bar pp\to\bar YY$
 which we use to fix the residual factor $C(t)$ in Eq.~(\ref{E1}).
 In Fig.~\ref{Fig:3} we show the differential cross section
 of the reaction $\bar pp\to\bar\Lambda\Lambda$ and
 $\bar pp\to\bar\Lambda\Sigma^0$
 as a function of the momentum transfer $t=(p_p-p_Y)^2$
 at the initial momentum
 $p_L=6$~GeV/c together with the available experimental
 data~\cite{Becker1978}.
The overall residual factor
\begin{eqnarray}
C(t)=\frac{0.37}{(1 - t/1.15)^2} \label{E16}
 \end{eqnarray}
provides a reasonable agreement of the calculation and the data.

\begin{figure}[ht]
    \includegraphics[width=0.35\columnwidth]{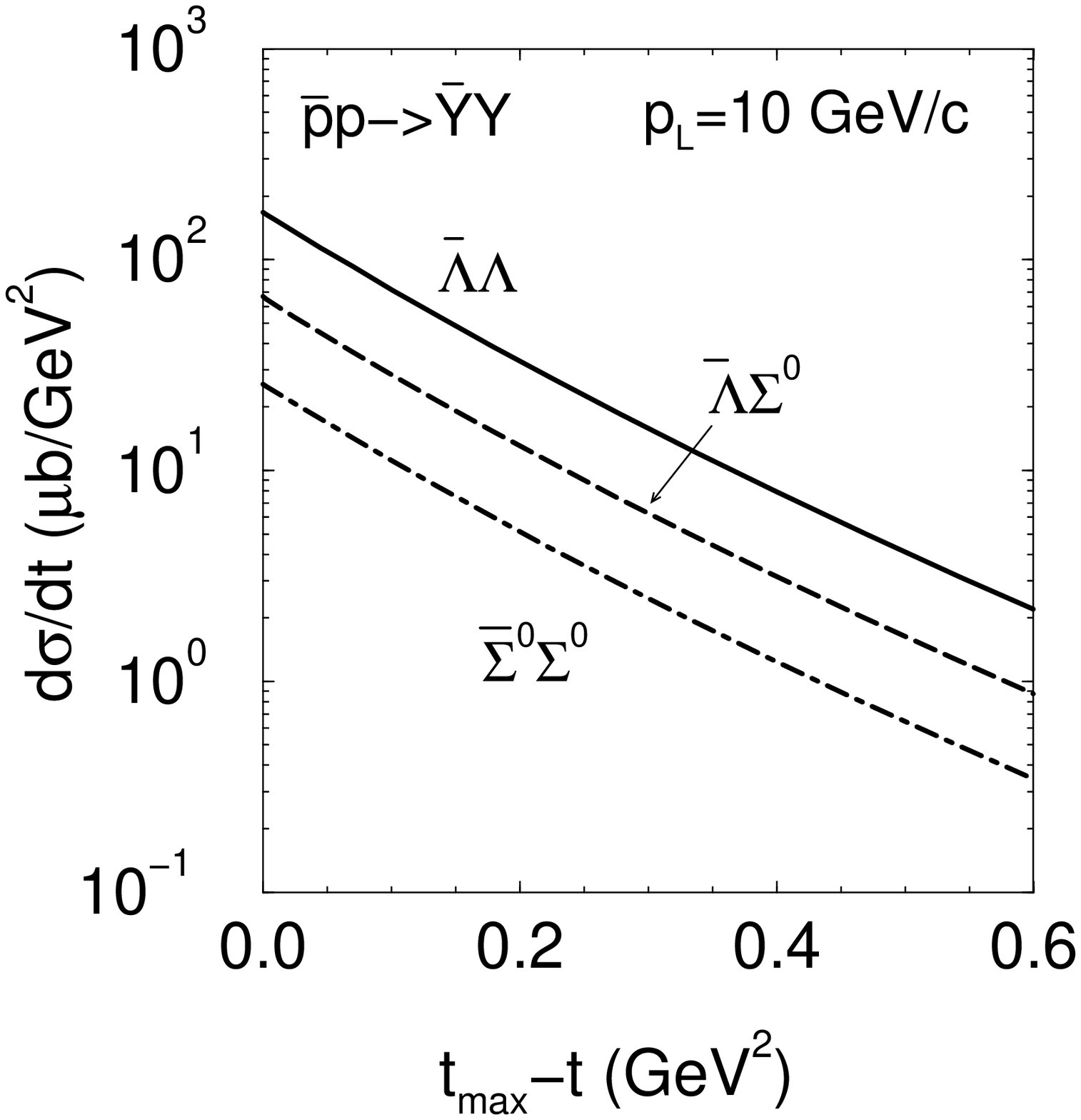}
    \qquad\qquad
    \includegraphics[width=0.35\columnwidth]{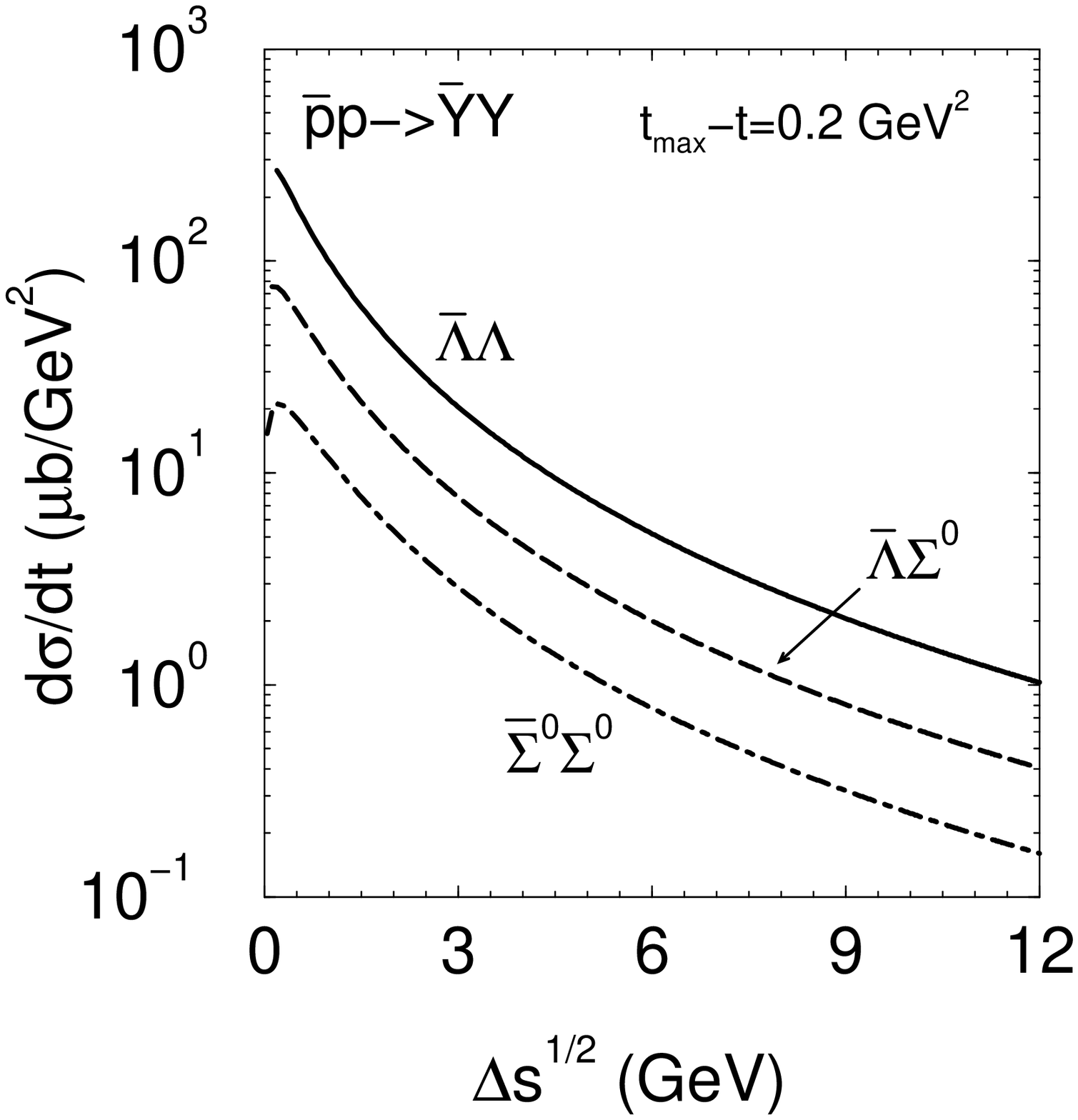}
   \caption{\small{Left panel:
   The differential cross sections of the reactions
   $\bar p p\to \bar \Lambda\Lambda$
   (solid curve), $\bar pp\to\bar\Lambda\Sigma^0$ (dashed curve),
 and $\bar pp\to\bar\Sigma^0\Sigma^0$
 (dot-dashed curve) as a function
 of $t_{\rm max}-t$ at $p_L=10$~GeV/c.
 Right panel:  The differential cross section as
 a function of the excess energy
 $\Delta s^{1/2}$ at $t_{\rm max}-t=0.2$~GeV$^2$.
\label{Fig:4}}}
  \end{figure}

 In Fig.~\ref{Fig:4} (left panel) we exhibit our prediction for
 the differential cross sections
 of the reactions $\bar p p\to\bar\Lambda\Lambda $,
 $\bar pp\to\bar\Lambda\Sigma^0\,(\bar\Sigma^0\Lambda)$ and
 $\bar pp\to\bar\Sigma^0\Sigma^0$
 as a function of $t_{\rm max}-t$ at initial momentum
 $p_L=10$~GeV/c. Here,
  $t_{\rm max}$ is the maximum momentum transfers which
  corresponds to the $\Lambda$  production at zero angle relative to
  the momentum of the incoming proton in the c.m.s.

 The exponential decrease of the cross section is defined
 by the Regge propagator $(s/s_i)^{2\alpha_{K^*}(t)}$ and the residual $C(t)$.
 The dependence on the excess energy
 $\Delta s^{1/2}\equiv \sqrt{s}-\sqrt{s_0}$, where
 $\sqrt{s_0}=M_{Y'} + M_{\bar Y}$, is shown in Fig.~\ref{Fig:4}
 (right panel). The calculation is done at fixed
  $ t_{\rm max}-t=0.2$~GeV$^2$.
 At large energies, the cross section behaves as $s^{2(\alpha_{K^*}-1)}
\simeq s^{-1.172}$. The ratio of the cross sections
with $\bar\Lambda\Lambda$, $\bar\Lambda\Sigma^0$ and
$\bar\Sigma^0\Sigma^0$ final states at large energy reads
\begin{eqnarray}
1 : r  :r^2,
\label{E17}
 \end{eqnarray}
where $r=(g_{K^*N\Lambda}/g_{K^*N\Sigma})^{-2}\simeq 0.4$.

   \begin{figure}[h!]
    \includegraphics[width=0.35\columnwidth]{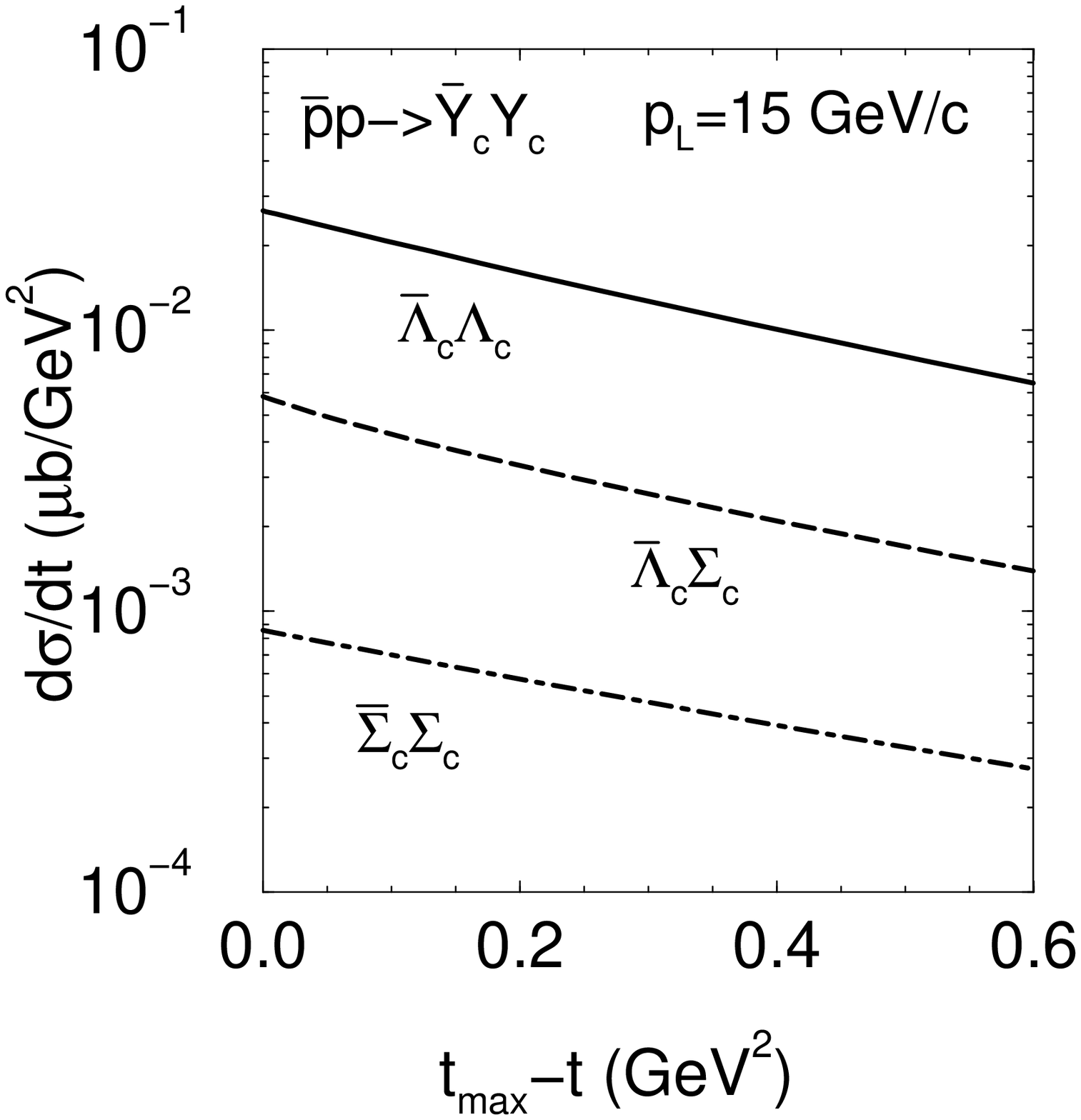}
    \qquad\qquad
    \includegraphics[width=0.35\columnwidth]{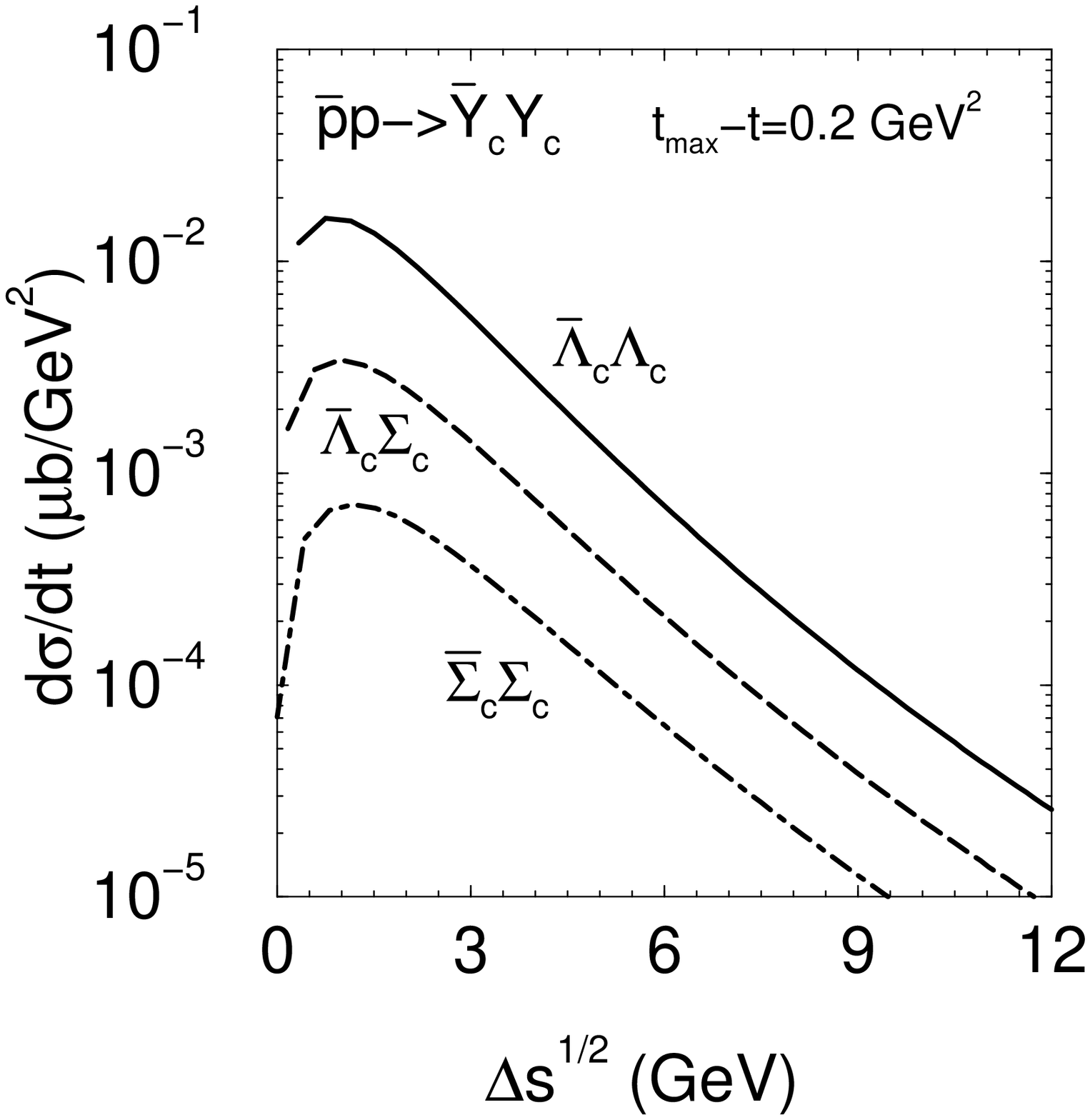}
   \caption{\small{Left panel:
   The differential cross sections of
   the reactions $\bar p p\to \bar \Lambda_c\Lambda_c$
   (solid curve), $\bar pp\to\bar\Lambda_c\Sigma_c$ (dashed curve),
  and $\bar pp\to\bar\Sigma_c\Sigma_c$(dot-dashed curve)
  as a function of $t_{\rm max}-t$ at $p_L=15$~GeV/c.
  Right panel: The  differential cross section
  as a function of the excess energy
  $\Delta s^{1/2}$ at $t_{\rm max}-t=0.2$~GeV$^2$.
  \label{Fig:5}}}
  \end{figure}
The predicted differential cross sections of the charm hyperon
production are exhibited in Fig.~\ref{Fig:5}. Here, we use the
notation $\Lambda_c\equiv\Lambda^+_c$ and
$\Sigma_c\equiv\Sigma^+_c$. The left panel shows the dependence on
$t_{\rm max} - t$ at fixed $p_L=15$~GeV/c. The right panel
exhibits dependence on the energy excess $\Delta s^{1/2}$ at fixed
$ t_{\rm max}-t=0.2$~GeV$^2$. The threshold initial momenta (for a
fixed target) for reactions with $\bar\Lambda_c\Lambda_c$,
$\bar\Lambda_c\Sigma_c$ and $\bar\Sigma_c\Sigma_c$ final states
are 10.15, 11.83, and 10.85 (GeV/c), respectively. The energy
excess at $p_L=15$~GeV/c for these final states are 0.90, 0.571
and 0.570 GeV, respectively. This energy is not asymptotically
high and some particular behavior of the cross sections in the
pre-asymptotical region is expected. Thus, in  Fig.~\ref{Fig:5}
(right panel) one can see a bump-like behavior at low $\Delta s$,
which reflects the energy dependence of $t_{\rm \max}$ in this
region.

\subsubsection{Longitudinal asymmetries}

For a better understanding of the results of our numerical calculation, it
seems to be useful to perform a qualitative analysis  of the
longitudinal asymmetry at forward production angle
(or $t=t_{\rm max}$), where the orbital interaction is absent. In this case, the
amplitude of the  $\bar pp\to\bar YY$ reaction may be
written as
\begin{eqnarray}
T_{m_f n_f;m_i,n_i}\sim R(s) \left(
A(s)\,\delta_{m_im_f}\,\delta_{n_in_f}
+\frac{1}{\sqrt{2}}B(s)\,(1-4m_im_f)\,\delta_{-m_im_f}\delta_{-n_in_f}\right)~,
\label{E18}
 \end{eqnarray}
where $R(s)$ is a spin-independent function, and $m_i,m_f,n_i$ and
$n_f$ stand for the spin projections of $p,Y,\bar p$,
and $\bar Y$, respectively, The longitudinal asymmetry is
expressed  through the spin-conserving ($A(s)$) and spin-flip
($B(s)$) amplitudes as
\begin{eqnarray}
{\cal A}=\frac{B^2(s)}{A^2(s)+B^2(s)}~. \label{E19}
 \end{eqnarray}
The spin-conserving  amplitude is determined by the two functions $a_0(s)$
and $a_\kappa(s)$
\begin{eqnarray}
 A(s)&=&(a_0(s) + a_\kappa(s))^2~,\nonumber\\
 a_0(s)&=& 1+\frac{{\bf p}_p {\bf p}_Y}
 {(E+M_N)(E+M_Y)}~,\nonumber\\
 a_\kappa(s)&=&\kappa\,a_0(s) -\frac{2\kappa E}{M_N+M_Y}
 \left( 1-\frac{{\bf p}_p {\bf p}_Y}
 {(E+M_N)(E+M_Y)}\right)~, \label{E20}
 \end{eqnarray}
where $\kappa$ is again the tensor coupling strength; ${\bf p}_p$, ${\bf
p}_Y$ denote the three momenta of the proton and the outgoing hyperon,
respectively, and $E=\sqrt{s}/2$ is the proton energy in c.m.s.
    \begin{figure}[ht!]
    \includegraphics[width=0.35\columnwidth]{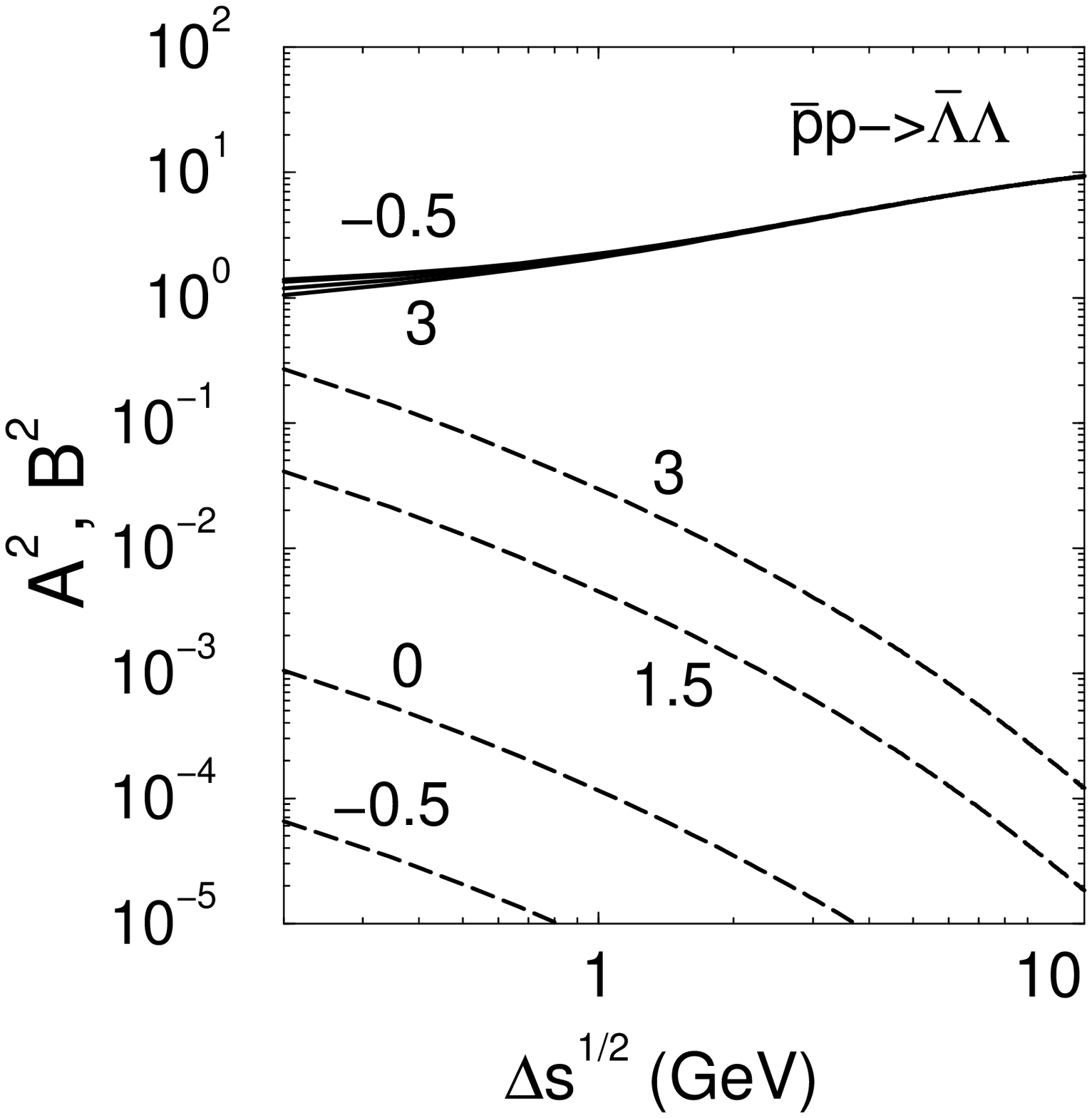}
    \qquad\qquad
    \includegraphics[width=0.35\columnwidth]{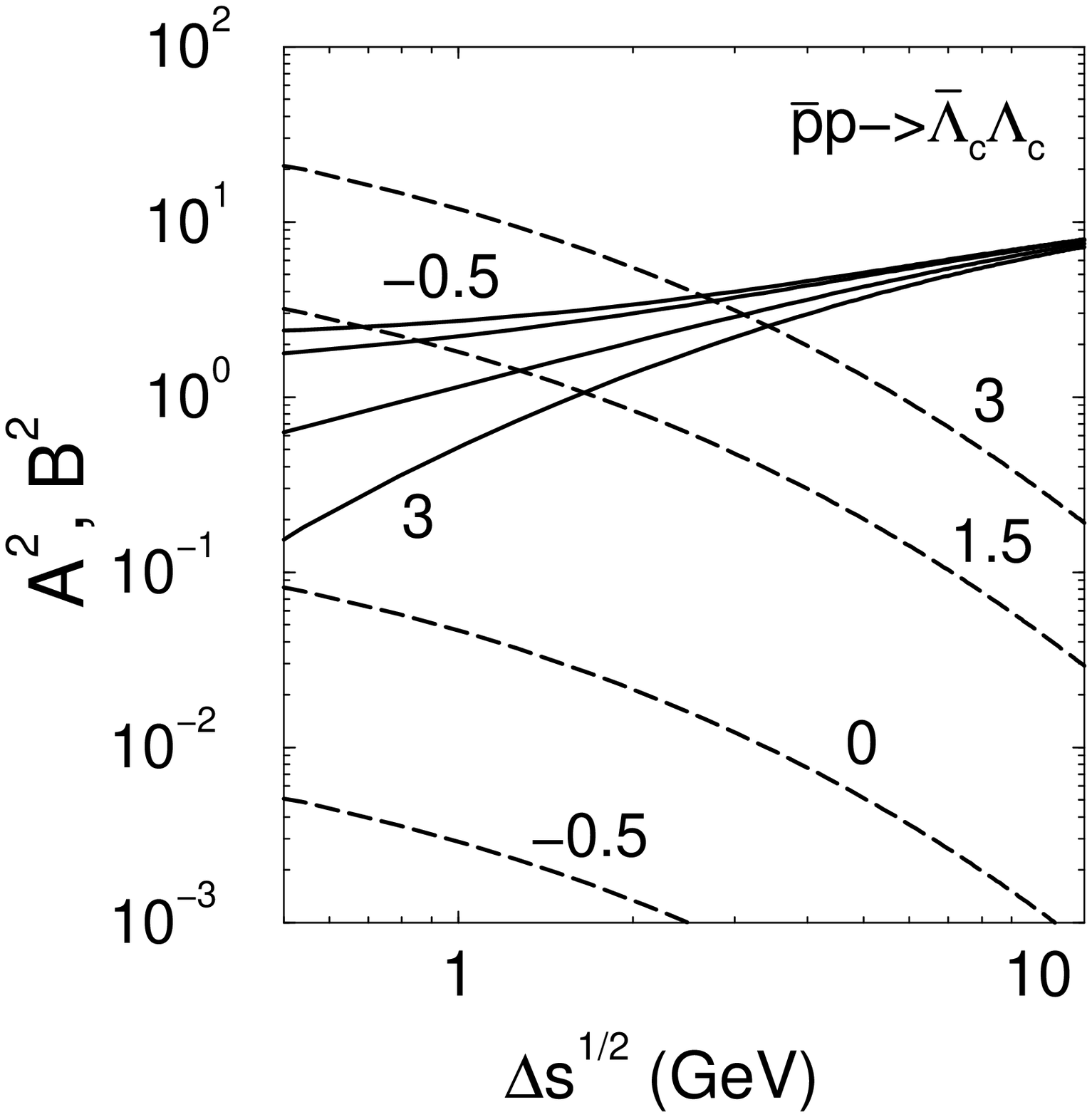}
   \caption{\small{
 The quantities $A^2$ (spin-conserving amplitudes, solid curves)
 and $B^2$ (spin-flip amplitudes, dashed curves) as a function of
 the excess energy $\Delta s^{1/2}$ with $t=t_{\rm max}$
 at different values of the
 tensor coupling $\kappa$. The left and right panels correspond to
 the reactions $\bar p p\to \bar \Lambda\Lambda$ and $\bar p p\to
 \bar \Lambda_c\Lambda_c$, respectively.
  \label{Fig:6}}}
  \end{figure}
In the case when $M_p\simeq M_Y$ or/and at high energies,
when $\sqrt{s}\gg M_Y$, $a_\kappa\to 0$
and the spin-conserving amplitude becomes independent of $\kappa$.

Contrarily, the spin-flip amplitude is proportional to the square of
the magnetic strength $(1+\kappa)^2$:
\begin{eqnarray}
B(s)=-\sqrt{2}\left((1+\kappa)( \frac{{\bf p}_p }{E+M_N}
- \frac{{\bf p}_Y }{E+M_Y})
\right)^2~.
\label{E21}
 \end{eqnarray}
At high energies with $E\gg M_Y$ and ${\bf p}_p\simeq {\bf p}_Y$,
$B(s)\to 0$ and, therefore, the asymmetry in Eq.~(\ref{E19})
vanishes. However, at finite energies and large values of
$(1+\kappa)$, the amplitudes $A(s)$ and $B(s)$ are comparable, and
the longitudinal asymmetry may be finite and large. The energy
dependence of $A^2(s)$ and $B^2(s)$ for the reactions $\bar pp\to
\bar \Lambda\Lambda$ and $\bar p p\to \bar \Lambda_c\Lambda_c$ is
shown in Fig.~\ref{Fig:6}.
In the $\bar \Lambda\Lambda$ final state,
the function $B^2(s)$ is rather small due to the small
difference between $M_N$ and $M_\Lambda$.
The dependence of $A^2(s)$ on the tensor coupling
$\kappa$ is rather weak. This leads to the small value of the
longitudinal asymmetry for the
$\bar p p\to \bar \Lambda\Lambda$ reaction.

In case of the $\bar p p\to \bar \Lambda_c\Lambda_c$ reaction, the
situation is quite different. The large difference between $M_N$ and
$M_{\Lambda_c}$ leads to a large value of $B^2(s)$, shown in
Fig.~\ref{Fig:6} (right panel), and results in a large value of the
longitudinal asymmetry.

For the $\bar pp\to \bar \Lambda\Sigma\,(\bar \Sigma\Sigma)$
reactions the spin-flip amplitude $|B(s)|$ is small because of
the small magnetic strength, $1+\kappa\simeq 0.09$, and the
asymmetry is almost zero.
    \begin{figure}[ht!]
    \includegraphics[width=0.35\columnwidth]{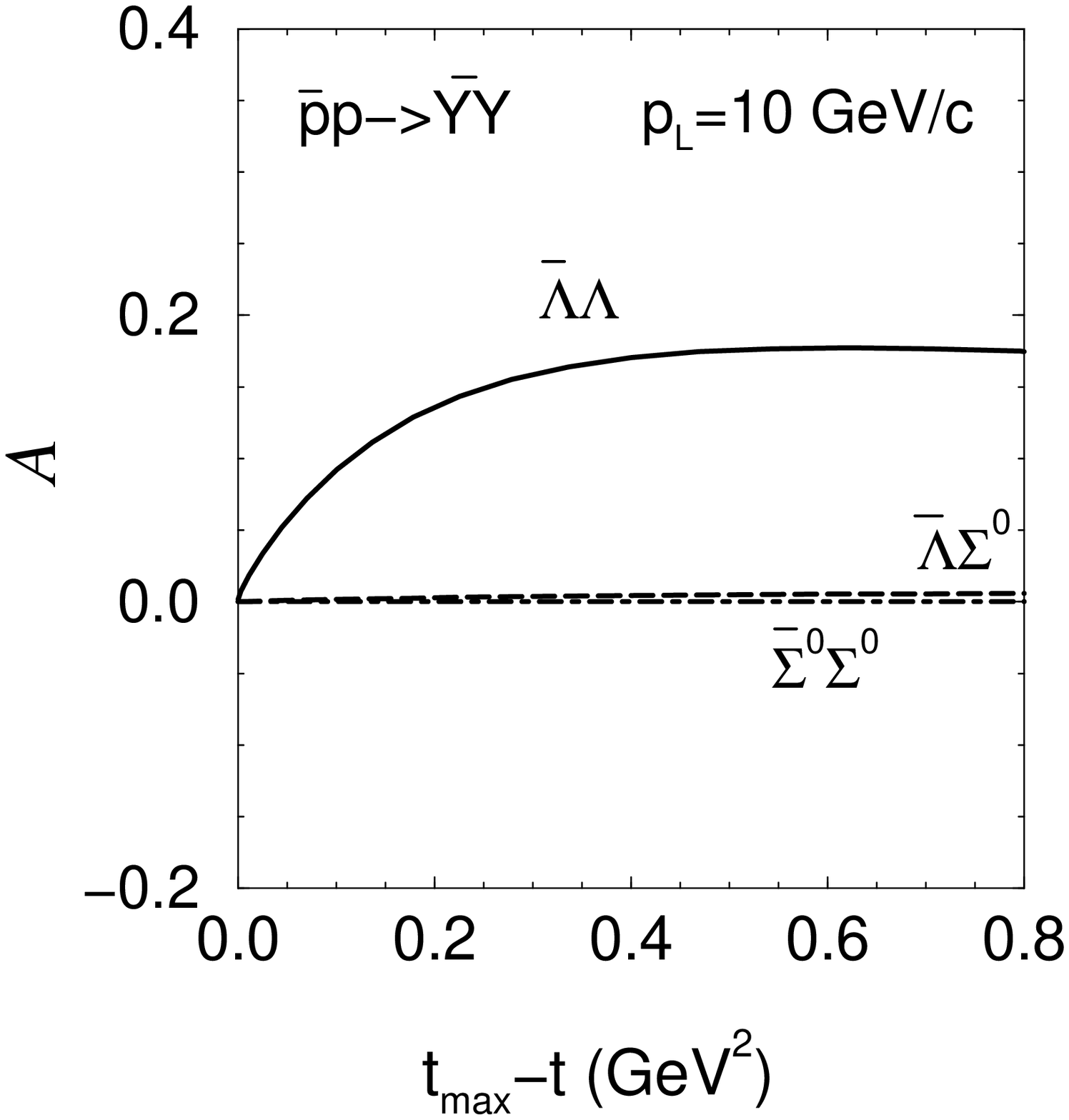}
    \qquad\qquad
    \includegraphics[width=0.35\columnwidth]{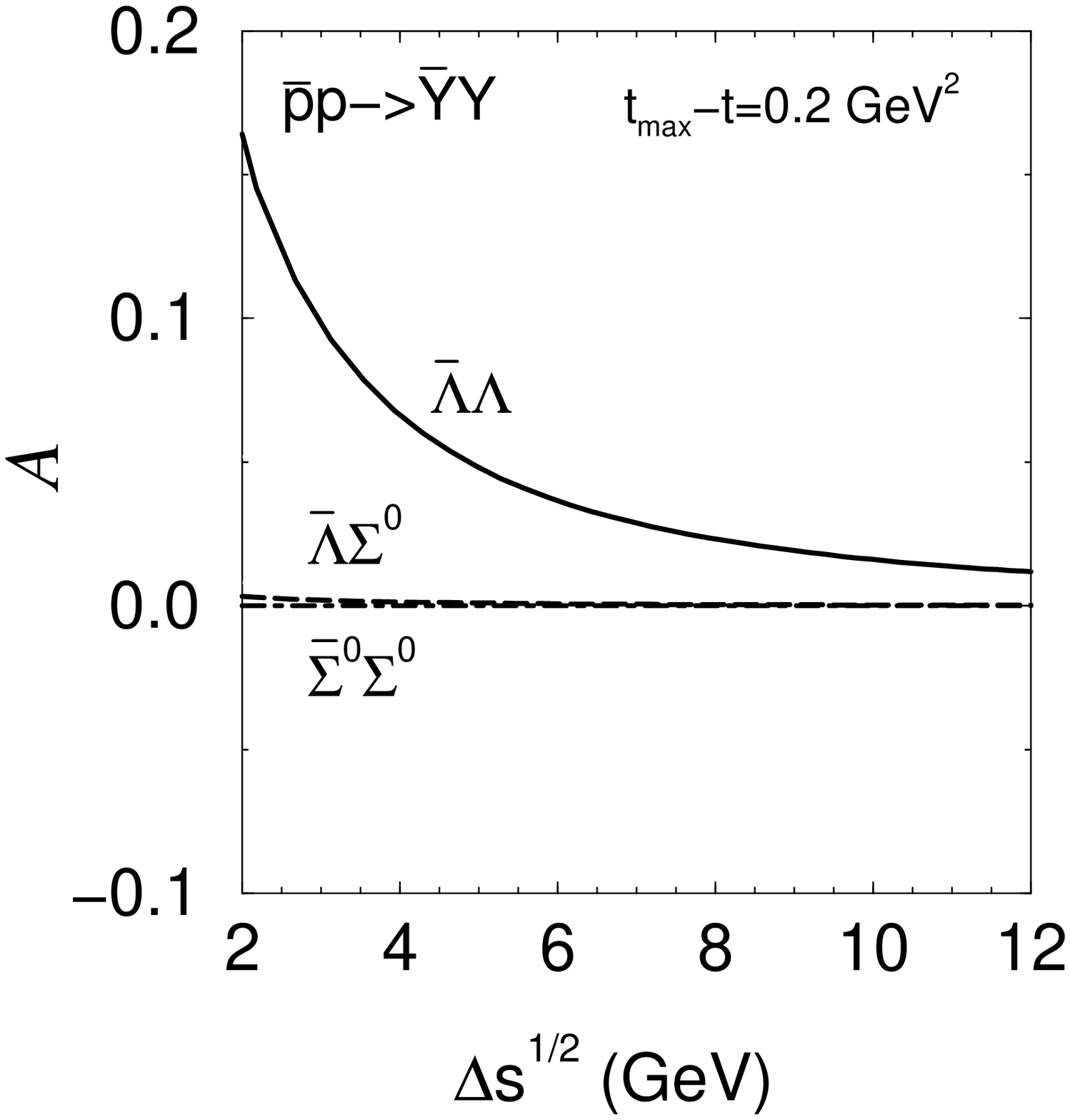}
   \caption{\small{Left panel:
   The longitudinal asymmetry for the reactions
   $\bar p p\to \bar \Lambda\Lambda$,
   $\bar\Lambda\Sigma^0$ and $\bar\Sigma^0\Sigma^0 $
   as a function of momentum transfer $t$ at $p_L=10$~GeV.
Right panel: The asymmetry  as a function of the excess energy
at $t_{\rm max}-t=0.2$~GeV$^2$.
\label{Fig:7}}}
  \end{figure}

    \begin{figure}[ht!]
    \includegraphics[width=0.39\columnwidth]{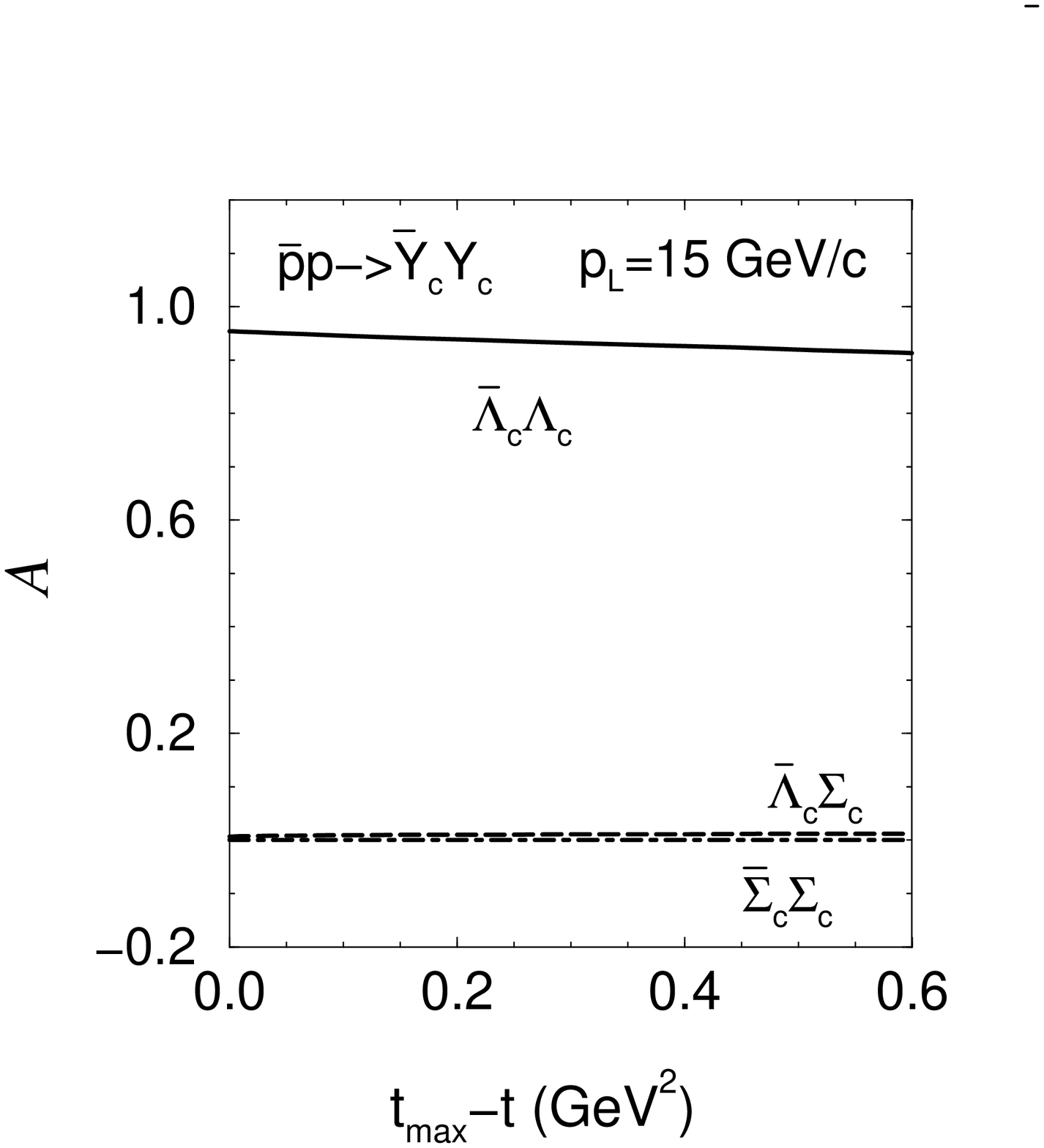}
    \qquad\qquad
    \includegraphics[width=0.35\columnwidth]{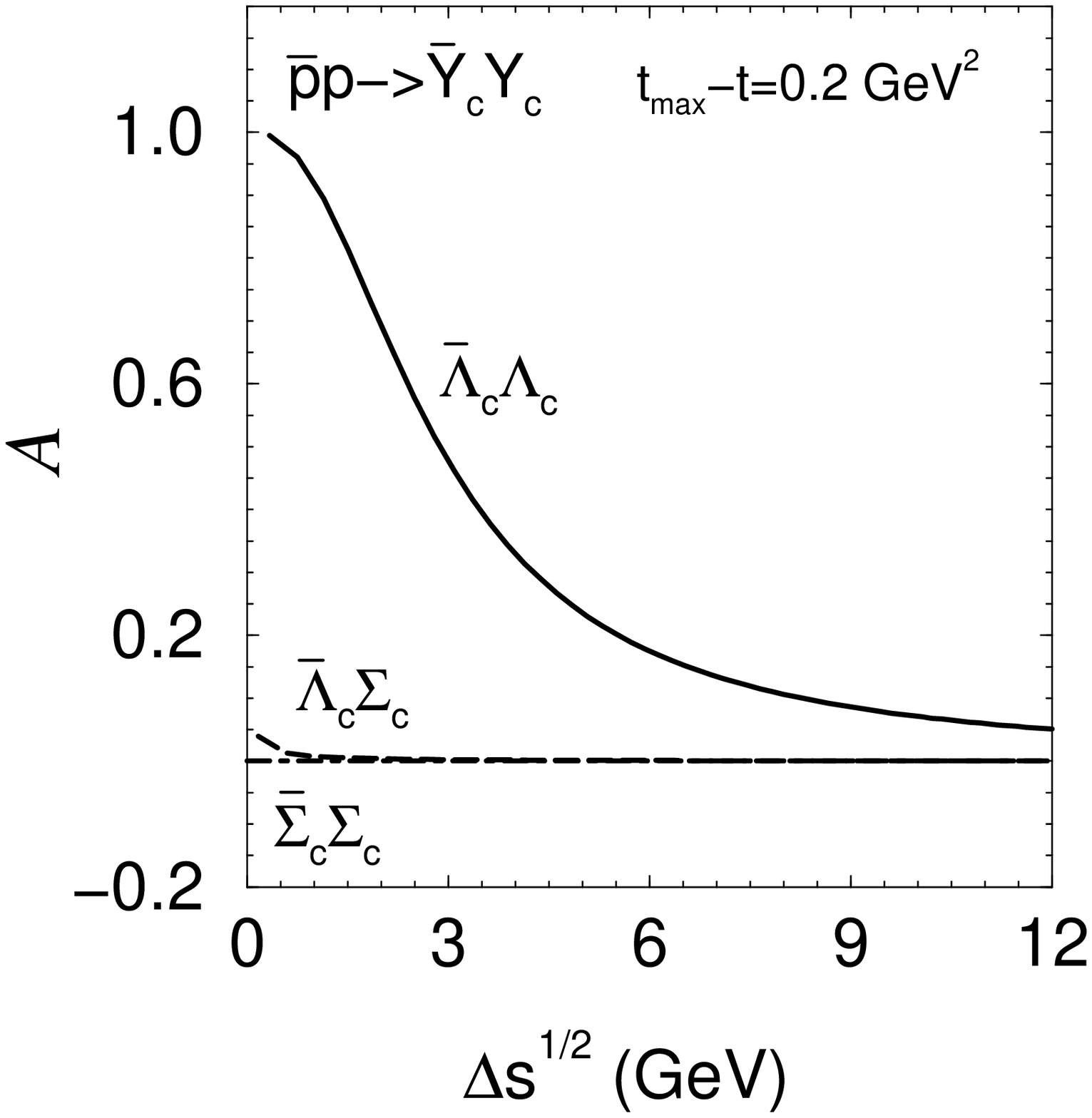}
   \caption{\small{Left panel:
   The longitudinal asymmetry  for
  the reactions $\bar p p\to \bar \Lambda\Lambda$,
   $\bar\Lambda\Sigma^0$ and $\bar\Lambda^0\Sigma^0 $
   as a function of momentum transfer $t$ at $p_L=10$~GeV.
Right panel: The asymmetry  as a function of the excess energy
at $t_{\rm max}-t=0.2$~GeV$^2$. \label{Fig:8}}}
  \end{figure}

Our predictions for the $\bar pp\to \bar YY$ reactions are shown
in Fig.~\ref{Fig:7}. The left panel exhibits the
$t$ dependence at the initial momentum $p_L = 10$~GeV/c. The right
panel shows the dependence on the energy excess at $t_{\rm
max}-t=0.2$~GeV$^2$. One can see that the result of the numerical
calculations agrees with our qualitative consideration. Thus, for
$\bar pp\to \bar\Lambda\Lambda$, the asymmetry  does not
exceed 0.2 at forward angles and decreases with energy. In the
$\bar pp\to \bar \Lambda\Sigma$ and $\bar \Sigma\Sigma$ reactions
it is almost zero.

The longitudinal asymmetry for the $\bar pp\to \bar Y_cY_c$
reactions is presented in Fig.~\ref{Fig:8}. In the case of the $\bar
pp\to \bar \Lambda_c\Lambda_c$ reaction, the asymmetry is large at
low energy excess and decreases rapidly with energy. In the reactions
$\bar pp\to \bar \Lambda_c\Sigma_c$ and $\bar \Sigma_c\Sigma_c$
the asymmetry is negligibly small.

\section{Reaction $\boldmath \bar pp \to \bar MM$ }

In this section, we discuss the production of $\bar MM$
(with $\bar MM$ being $\bar KK$ or $D\bar D$)
in $\bar pp$ collisions. We assume that at small momentum transfer $-t$,
where
$t=(p_p-p_K)^2$ or $t=(p_p-p_{\bar D})^2$, the dominant contribution comes
from the baryon exchange channels.

\begin{figure}[ht]
   \includegraphics[width=0.7\columnwidth]{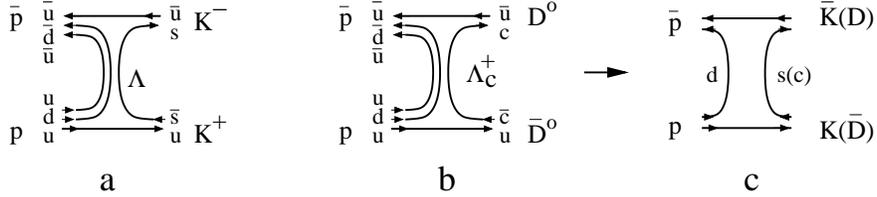}
   \caption{\small{Planar diagrams for the reaction  $\bar p p \to K^-K^+$ (a)
  and $\bar p p \to D^0\bar D^0$ (b). The exchanged baryon as
  a quark-diquark object (c). The symbol $d$ stands for a $qq$ diquark.}}
   \label{Fig:9}
   \end{figure}

As an example, in Fig.~\ref{Fig:9} (a) and (b) we show the planar
diagrams for $\bar pp\to K^-K^+$ and
 $\bar pp\to  D^0\bar D^0$, with $\Lambda$ and $\Lambda_c^+$
exchange, respectively. The cases of $\Sigma$ ($\Sigma_c^+$)
exchange, or  $\Sigma^+$ ($\Sigma_c^{++}$) exchange for $\bar KK$
($D^+D^-$) final state, are similar. Here and further on we employ
the quark-diquark identity, used in many phenomenological
approaches to QCD~\cite{Close_TB,FS,BurakovskyGoldmanHorvitz}.
This means that the exchanged baryon is considered as a
quark-diquark string object, shown schematically in
Fig.~\ref{Fig:9} (c).

\subsection{Reaction
$\boldmath \bar pp\to \bar KK $}

Let us consider first the reaction $\bar pp\to \bar KK$.
For definiteness, we consider $\bar pp\to K^-K^+$ with
$\Lambda$ exchange. The cases of $\Sigma$ and $\Sigma^+$ for
$\bar K^0K^0$ final states may be executed in an analog way.

The assumption of the quark-diquark identity allows to generalize
our model developed in the previous section. Namely,
we assume that the amplitude
of the reaction $\bar pp\to K^-K^+$ has the form of a Regge
pole amplitude dominated by the $\Lambda+\Sigma$ exchange
trajectories. Thus, for the $\Lambda$ exchange it reads
\begin{eqnarray}
T^{\bar pp\to K^-K^+}_{m_i,n_i} =
C'(t){\cal M}^{\bar pp\to K^-K^+}_{m_i,n_i}(s,t)
\frac{g_{KN\Lambda}^2}{s_0}\, \Gamma(\frac{1}{2} - \alpha_{
ds}(t))\, \left(-\frac{s}{s_{\bar pp:\bar KK}} \right)^{\alpha_{
ds}(t)-\frac12},
\label{E22}
\end{eqnarray}
where $m_i$ and $n_i$ are the spin projections of $p$ and $\bar p$,
respectively, $d$ stands for a $ud$ diquark, $\alpha_{ds}(t)$
is the $\Lambda$ trajectory, $g_{KN\Lambda}^2$ is the coupling
constant of the $KN\Lambda$ interaction and $s_0=1$~GeV is an
universal scale parameter. The spin dependence is accumulated in
the amplitude ${\cal M}$ which in the limit of $s\to \infty$
results in ${\cal M} \approx \sqrt{s} $. The explicit form of
${\cal M}$ will be defined later. The overall residual function $C'(t)$
will be found again from a comparison with available experimental
data.

The parameters of the trajectory for the non-diagonal transition
$\alpha_{ ds}$ are related to the corresponding parameters for the
"diagonal" transitions $\alpha_{\bar ss}$ and $\alpha_{\bar dd}$
similarly to Eqs.~(\ref{E2}) and (\ref{E3})
\begin{eqnarray}
2\alpha_{ ds}(0)&=&\alpha_{\bar dd}(0)+\alpha_{\bar ss}(0)~,
\label{E23}\\
{2}/{\alpha_{ds}'}&=& 1/{\alpha_{\bar dd}'}+{1}/{
\alpha_{\bar ss}'}~. \label{E24}
 \end{eqnarray}
Using the $\Lambda$ trajectory as input~\cite{Storrow_PR}
\begin{eqnarray}
\alpha_{ds}=\alpha_\Lambda=-0.65+0.94t~, \label{E25}
\end{eqnarray}
and $\alpha_{\bar ss}$ from Eq.~(\ref{E6}), one can evaluate the
diagonal $\alpha_{\bar dd}$ trajectory at small $|t|$ as
\begin{eqnarray}
\alpha_{\bar dd}(t)=-1.58 +  \alpha'_{\bar dd}\,t
\label{E26}
 \end{eqnarray}
with $ \alpha'_{\bar dd}=1.542$~GeV$^{-2}$.

The equation for the energy scale parameter $s_{\bar pp:\bar KK}$
is slightly different from Eq.~(\ref{E7}). Now it reads
 \begin{eqnarray}
 \left(s_{\bar pp:\bar KK}\right)^{2(\alpha_{ds}(0)-\frac{1}{2})}
 = \left(s_{\bar pp}\right)^{\alpha_{\bar dd}(0)}\times
 \left(s_{\bar KK}\right)^{\alpha_{\bar ss}(0)-1}~.
 \label{E27}
 \end{eqnarray}
Using $s_{\bar KK}=1.21$~GeV$^2$ and $s_{\bar pp}=2.25$~GeV$^2$,
one gets $s_{\bar pp:\bar KK}=1.853$~GeV$^2$.

The spin dependence of the amplitude in Eq.~(\ref{E22}) is
determined by the form of the $KN\Lambda$ interaction given by the
effective Lagrangian
\begin{eqnarray}
 {\cal L}_{NYK}
 =  - i\bar{N}\,\gamma_5\, YK +{\rm h.c.}~,
\label{E28}
\end{eqnarray}
where $N,Y$ and $K$ denote the nucleon, hyperon and the $K$ meson
fields, respectively, $Y$ stands for $\Lambda$, $\Sigma$, and so
on. This form leads to the following expression of the amplitude
${\cal M}$ in Eq.~(\ref{E22})
\begin{eqnarray}
 &&{\cal M}^{\bar pp\to\bar KK}_{m_i n_i}(s,t)= {\cal
 N}(s,t)\,\left[\bar v_{n_i}\,
 (\fs p_Y - M_Y )\, u_{m_i}\right]~,\nonumber\\
 && {\cal N}(s,t)=
\frac{F_{\infty}(s)}{F(s,t)},\qquad F^2_{\infty}(s)=s\,M_Y^2/2~,
 \nonumber\\
 &&F^2(s,t) =\frac12\,\left( (s-2M_N^2)(M_Y^2-t)
 +4M_NM_Y(M_N^2+M_K^2+t \right.
 \nonumber\\
 &&\left. -(M_N^2-M_K^2+t)^2 -M_N^2(M_Y^2+t)
\right)~, \label{E29}
\end{eqnarray}
where $p_Y=p_p-p_{K}$.

For the $KNY$ coupling constants we use the average values of the
Nijmegen potential~\cite{Stoks1999}: $g_{KNY}=-15.755$, for
$Y=\Lambda$ and $-4.785$ for $Y=\Sigma^0$.

In case of the $\bar pp\to K^-K^+$ reaction, the total amplitude is
the coherent sum of the $\Lambda$ and $\Sigma$ exchange
trajectories, while the reaction $\bar pp\to \bar K^0K^0$ is dominated by
the $\Sigma^+$ trajectory. Following~\cite{Storrow_PR} we use
\begin{eqnarray}
\alpha_{\Sigma}\simeq\alpha_{\Sigma^+}
\simeq-0.79+0.87t.
\label{E30}
\end{eqnarray}
For simplicity, for the $\Sigma$ exchange channels
we use the same the energy scale parameter as
for the $\Lambda$ exchange, taking into account similarity of the
corresponding trajectories and the fact that
the contribution of $\Sigma$ exchange amplitude is much smaller
than the dominant  $\Lambda$ exchange one.

\subsection{Reaction
$\boldmath \bar pp\to D\bar D $}

In this case, the amplitude is defined by Eq.~(\ref{E22}) with the
substitutions $\Lambda\to \Lambda^+_c\equiv\Lambda_c$,
$\Sigma^0\to\Sigma_c^+$, $\Sigma^+\to\Sigma_c^{++}$ $K^+\to \bar
D^0$, $K^-\to D^0$, $K^0\to D^-$ and $\bar K^0\to D^+$ and  so on.
As above, we assume the validity of SU(4) symmetry which means
that the coupling constants of the $DNY_c$ interaction are chosen
to be the same as for the case of $KNY$ interaction. The
$\Lambda_c$ trajectory is calculated using
\begin{eqnarray}
2\alpha_{ dc}(0)&=&\alpha_{\bar dd}(0)+\alpha_{\bar cc}(0)~,
\label{E31}\\
{2}/{\alpha'_{ dc}}&=& {1}/{\alpha_{\bar dd}'}+{1}/{ \alpha_{\bar
cc}'}~, \label{E32}
 \end{eqnarray}
where $\alpha_{\bar cc}(t)\equiv\alpha_{J/\psi}(t)$ and $\alpha_{
dd}(t)$ are defined by Eqs.~(\ref{E15}) and (\ref{E26}),
respectively. Thus, for $\alpha_{dc}(t)$ and the energy scale
parameter
 $s_{\bar pp:D\bar D}$ we have
\begin{eqnarray}
&&\alpha'_{dc}(0)\simeq-2.09,\qquad \alpha'_{ dc}\simeq
0.557\,{\rm GeV}^{-2}~,
\label{E33}\\
&&s_{\bar pp:D\bar D}\simeq 3.59\,{\rm GeV}^{2}~.
\label{E34}
\end{eqnarray}
For simplicity, we assume
$\alpha_{\Lambda^+_c}\equiv\alpha_{dc}
\simeq\alpha_{\Sigma^+_c}
\simeq\alpha_{\Sigma^{++}_c}$.

\subsection{Results}

\subsubsection{Differential cross sections}
\begin{figure}[ht!]
    \parbox{.5\textwidth}{
    \includegraphics[width=0.36\columnwidth]{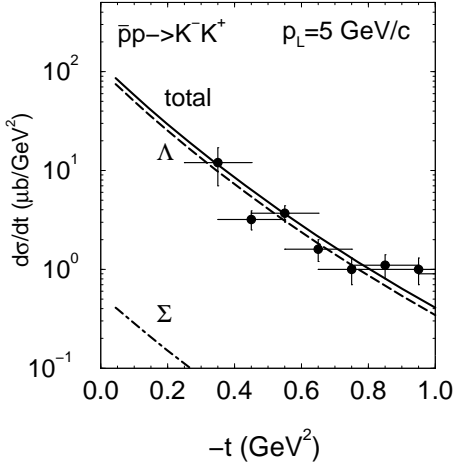}
}
\hfill
\parbox{.4\textwidth}{
   \caption{\small{
   Differential cross section of the $\bar p p\to K^-K^+$
   reaction as a function of momentum transfer $t$ at $p_L=5$~GeV.
   The contributions from $\Lambda$ and $\Sigma$ exchanges are shown
   by dashed and dot dashed curves, respectively.
  The experimental data are taken from  Ref.~\protect\cite{Eide_NP}.
\label{Fig:10} }}}
  \end{figure}

 The differential cross section
 of the $\bar pp\to\bar K^-K^+$ reaction
 as a function of the momentum transfer $t=(p_p-p_{K^+})^2$
 at initial momentum
 $p_L=5$~GeV/c together with available experimental
 data \cite{Eide_NP} is presented in Fig.~\ref{Fig:10}.
 The separate contributions from $\Lambda$ and $\Sigma^0$
 exchange are shown
 by dashed and dot dashed curves, respectively. The solid curve is the
 coherent sum of these contributions. One can see a dominance
 of the $\Lambda$  exchange trajectory in $\bar pp\to\bar K^-K^+$.
 This reaction
 is used to fix the residual factor $C'(t)$ in Eq.~(\ref{E22}).
 We find
\begin{eqnarray}
 C'(t)=\frac{0.52}{(1 - t/1.15)^2}~,
 \label{E35}
 \end{eqnarray}
 i.e., it  coincides within $\sim 30$\% with the residual
 in $\bar pp\to \bar YY$ reactions (cf.\ Eq.~(\ref{E16})),
 which is in favor of the consistency of the model.

\begin{figure}[ht]
    \includegraphics[width=0.35\columnwidth]{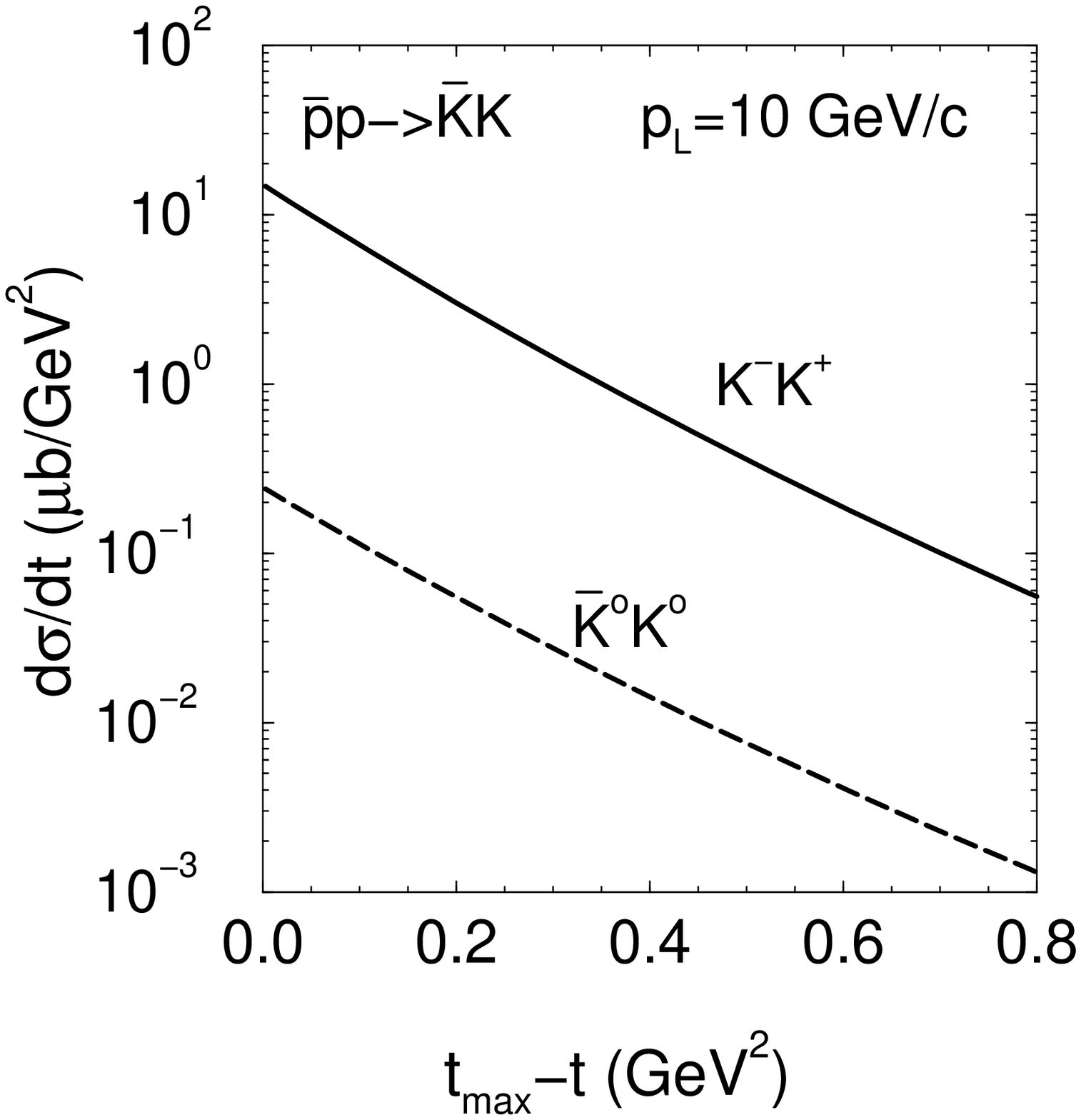}
    \qquad\qquad
    \includegraphics[width=0.35\columnwidth]{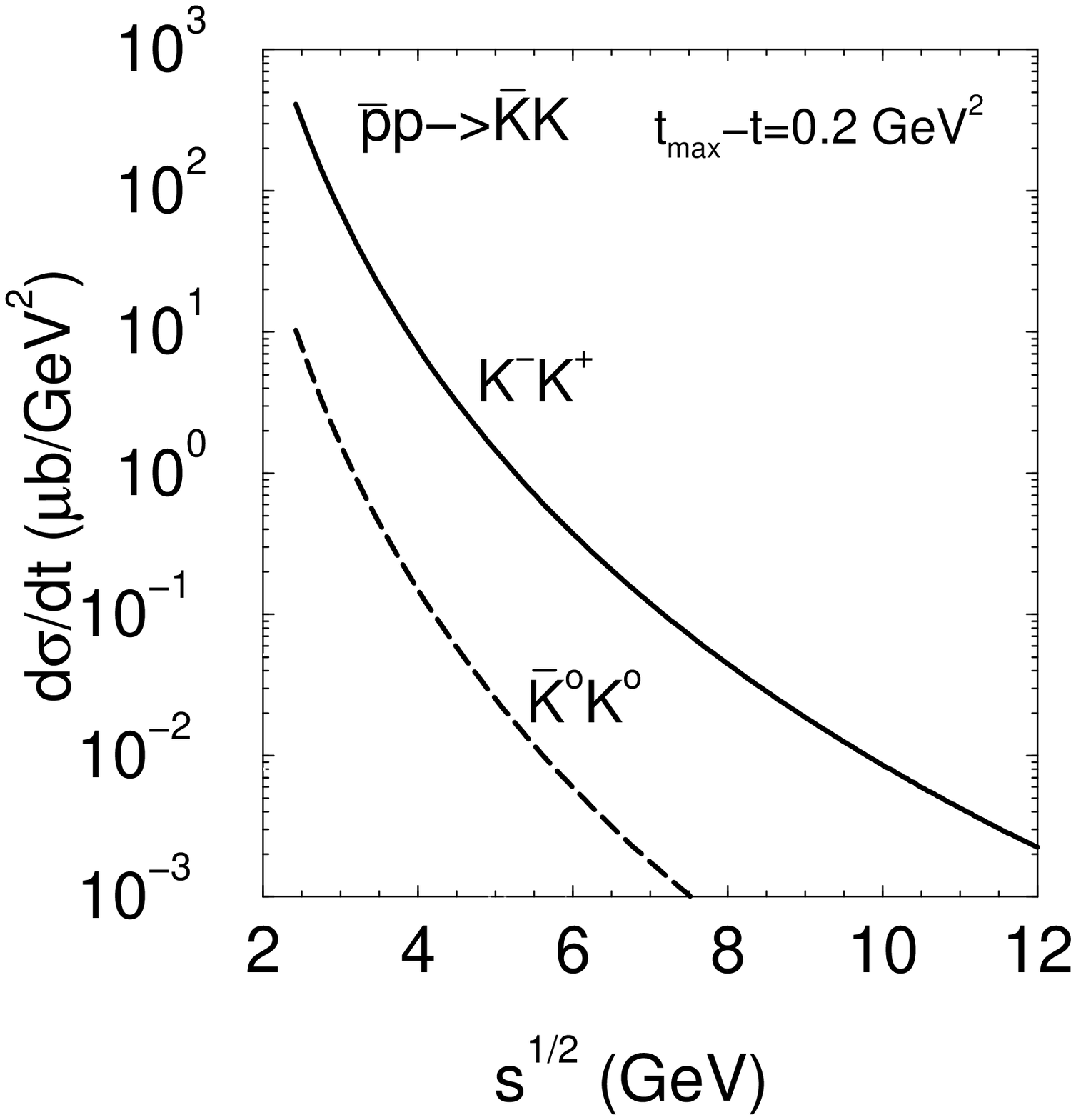}
   \caption{\small{Left panel:
   The differential cross sections
   of the reactions $\bar p p\to K^-K^+$
   (solid curve) and $\bar pp\to\bar K^0K^)$ (dashed curve)
   as a function
  of $t_{\rm max}-t$ at $p_L=10$~GeV/c.
 Right panel: The differential cross sections
  as a function of the energy
  $\sqrt{s}$ at $t_{\rm max}-t=0.2$~GeV$^2$.
\label{Fig:11}}}
  \end{figure}

 In Fig.~\ref{Fig:11} (left panel), we show our prediction for
 the differential cross sections
 of the reactions $\bar p p\to  K^-K^+$,
 and $\bar p p\to \bar K^0K^0$
 as a function of $t_{\rm max}-t$ at the initial momentum
 $p_L=10$~GeV/c.
 The dependence  of the differential cross sections
 on energy ($\sqrt{s}$) at fixed  $ t_{\rm max}-t=0.2$~GeV$^2$,
 is exhibited in Fig.~\ref{Fig:11}
 (right panel).
 At large energies the cross sections behave
 as $\sim s^{-2.3}$ and $\sim s^{-3.58}$ for
 the  $K^-K^+$ and $\bar K^0K^0$ final states, respectively.

\begin{figure}[h!]
    \includegraphics[width=0.35\columnwidth]{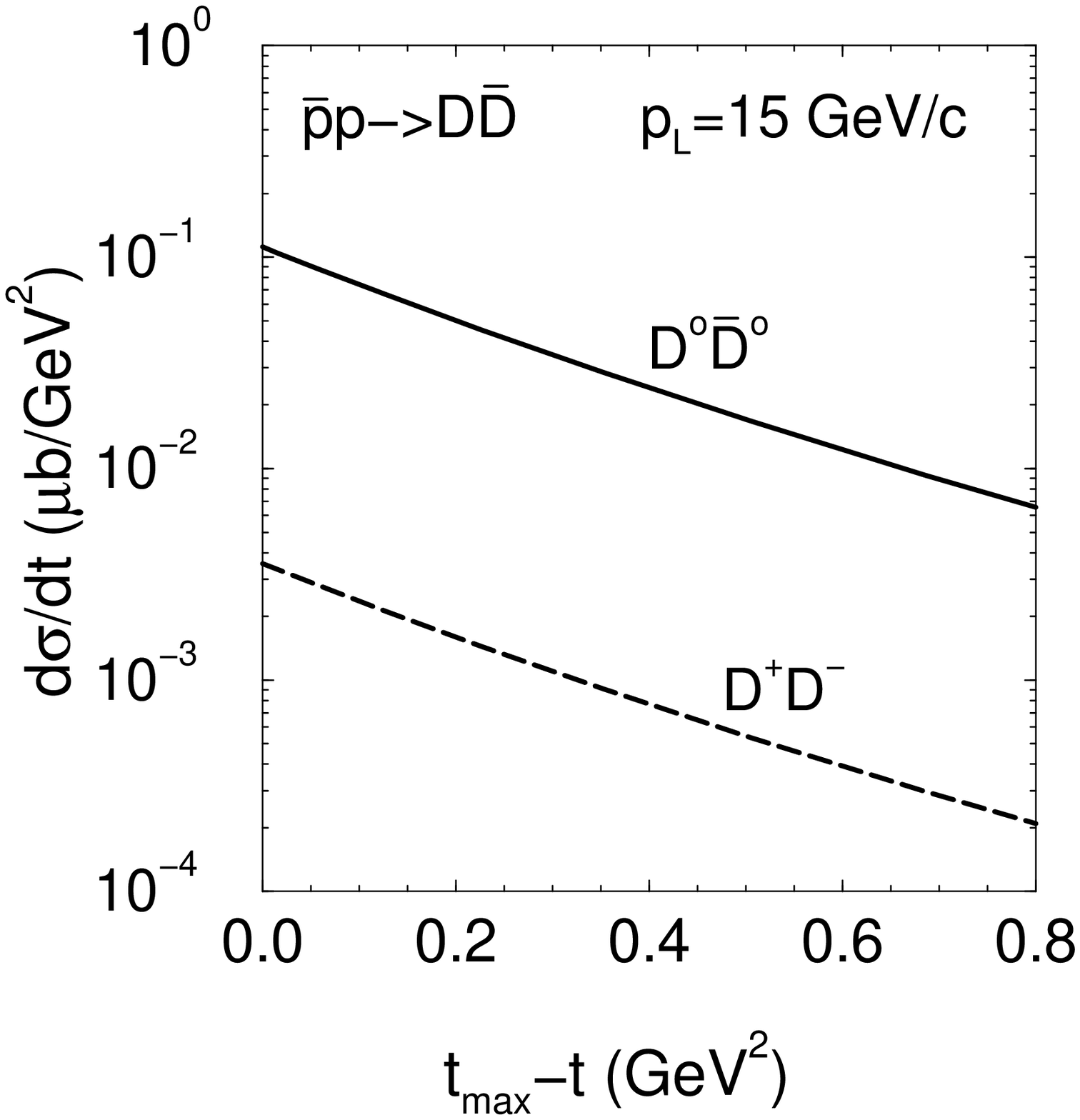}
    \qquad\qquad
    \includegraphics[width=0.35\columnwidth]{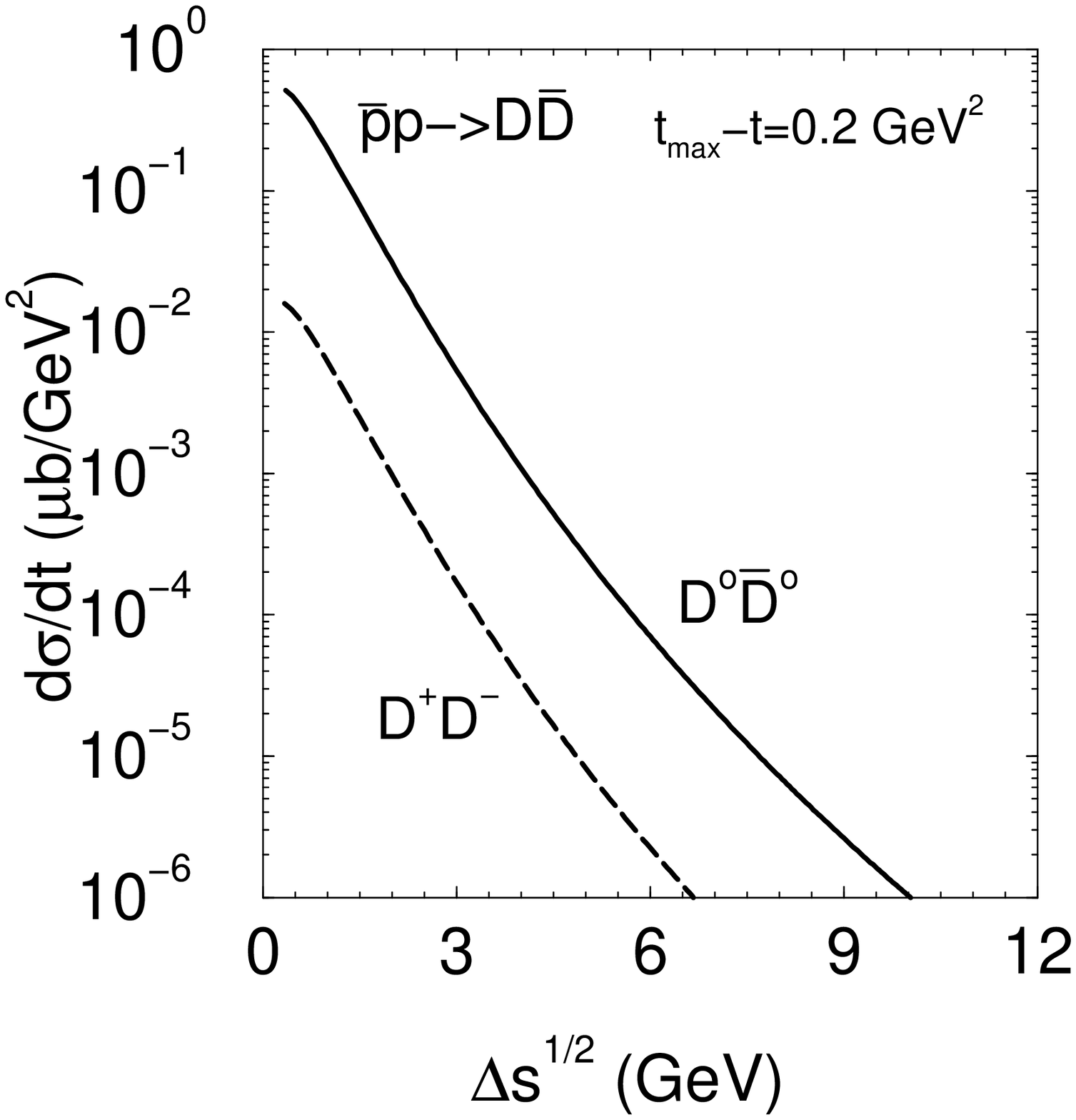}
   \caption{\small{Left panel:
  The differential cross sections of the reactions
  $\bar p p\to \bar D^0D^0$
   (solid curve) and $\bar pp\to D^-D^+$ (dashed curve),
 as a function of $t_{\rm max-t}$ at $p_L=15$~GeV/c.
 Right panel:  Differential cross section as
 a function of the excess energy
 $\Delta s^{1/2}$ at $t_{\rm max}-t=0.2$~GeV$^2$.
\label{Fig:12}}}
  \end{figure}

Our prediction for the differential cross sections
of $D\bar D$ pair production is presented in Fig.~\ref{Fig:12}.
The left panel illustrates the dependence on $t_{\rm max} - t$ at
fixed $p_L=15$~GeV/c. The right panel exhibits the
dependence on the excess energy $\Delta s^{1/2}$ at fixed
$ t_{\rm max}-t=0.2$~GeV$^2$. The ratio of the cross
sections with $D^-D^+$ and   $\bar D^0D^0$ final states
is close to
$(\sqrt{2}\,g_{KN\Sigma}/g_{KN\Lambda})^4\simeq0.034$.
The cross sections decrease rapidly with energy as $s^{-6.18}$,
therefore, the region with small excess energy is more
suitable for studying these reactions.

\subsubsection{Longitudinal asymmetry}

In reactions $\bar pp\to\bar KK$ ($D\bar D$) at forward production
angle (or $t=t_{\rm max}$), the spin in the final state is equal
to zero. This means that the production amplitude may be expressed
as
\begin{eqnarray}
T_{m_i,n_i}\sim B(s)\,\delta_{m_i-n_i}~,
\label{E36}
 \end{eqnarray}
and therefore, the asymmetry in Eq.~(\ref{E14}) ${\cal A}=1$. At
finite angles, the spin-orbital interactions becomes important,
which leads to an increase of the contribution of
$d\sigma^\rightrightarrows$ to the total cross section and
to a decrease of the asymmetry.

    \begin{figure}[ht!]
    \includegraphics[width=0.35\columnwidth]{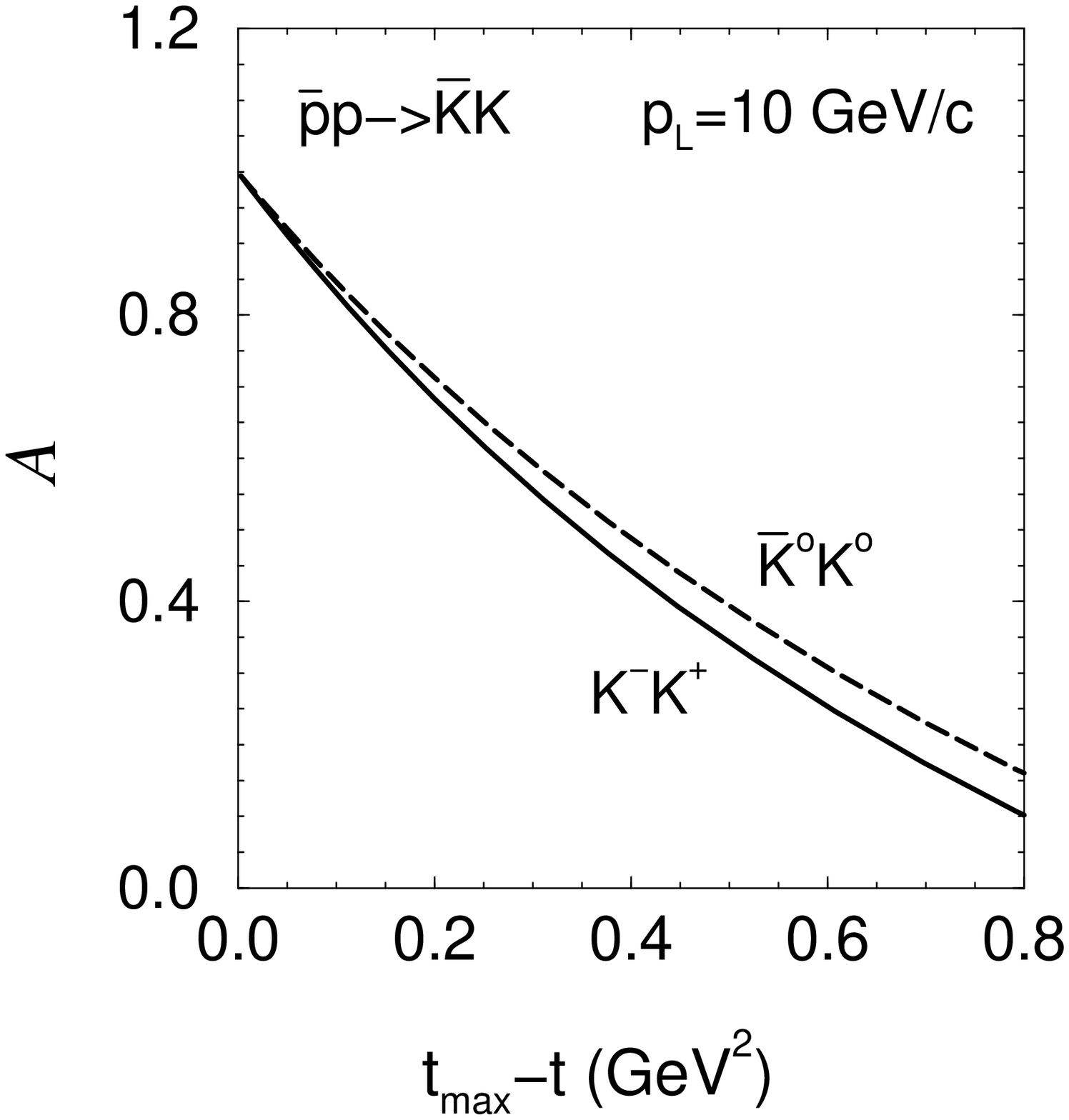}
    \qquad\qquad
    \includegraphics[width=0.35\columnwidth]{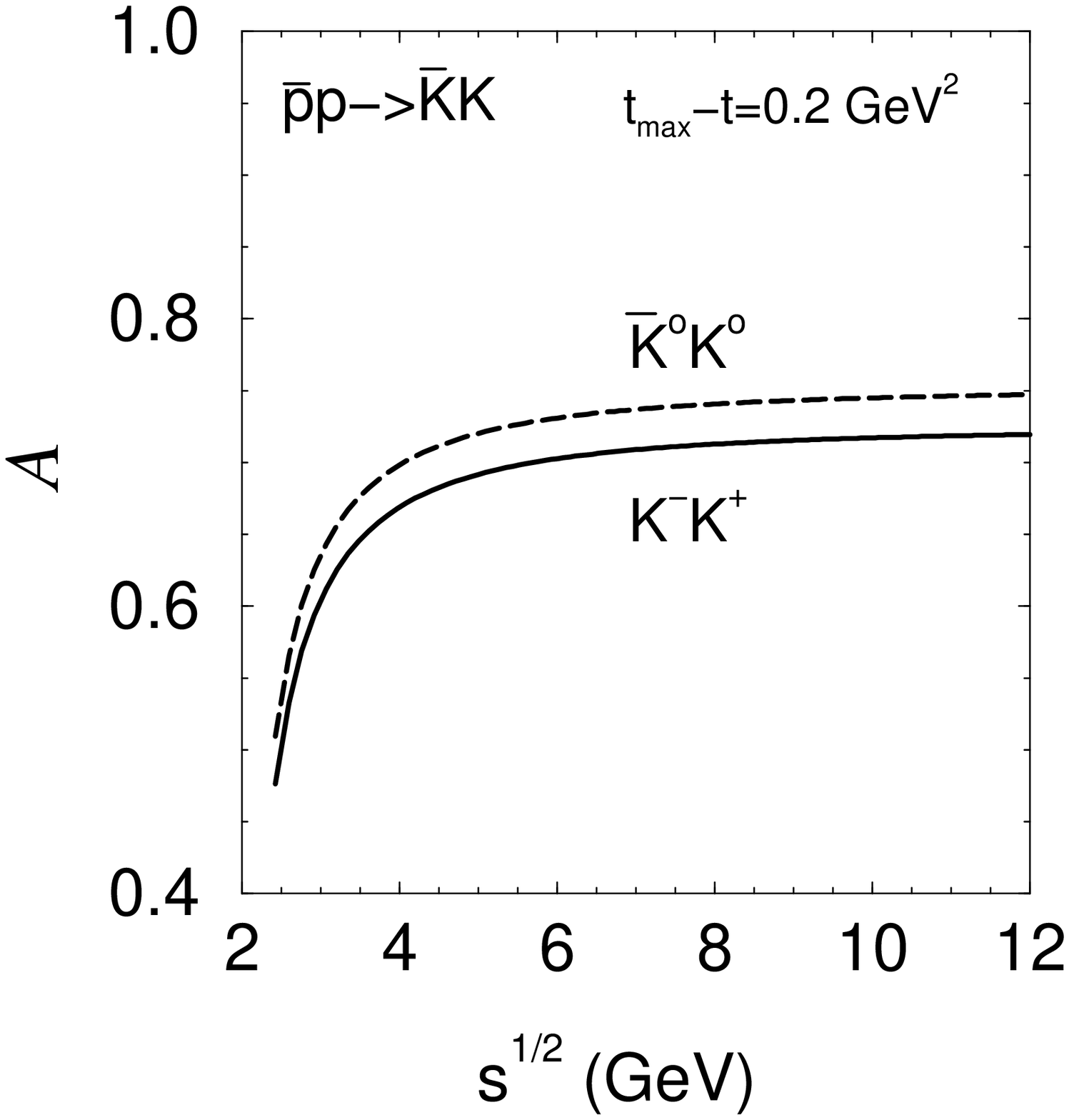}
   \caption{\small{
   The longitudinal asymmetry  for reactions
   the $\bar p p\to K^-K^+$ (solid curves)
   and  $\bar p p\to \bar K^0K^0$, (dashed curves).
   Left panel: The asymmetry
   as a function of momentum transfer
   $t_{\rm max}-t$ at $p_L=10$~GeV.
   Right panel:  The asymmetry  as a function of the energy
    $\sqrt{s}$ at $t_{\rm max}-t=0.2$~GeV$^2$.
\label{Fig:13}}}
  \end{figure}

In Fig.~\ref{Fig:13} we show our prediction for
$\bar pp\to\bar KK$. The left panel exhibits the $t$
dependence at $p_L = 10$~GeV/c, whereas the right panel exhibits
the $\sqrt{s}$ dependence at $t_{\rm max}-t=0.2$~GeV$^2$. One can
see a decrease of ${\cal A}$ with $-t$ and an almost constant
value at large $\sqrt{s}$ and fixed $t_{\rm max}-t$. Some
difference in ${\cal A}$ for $\bar K^0K^0$ and $K^+K^-$ final
states is explained by the difference of masses of $K^0$ and
$K^\pm$ mesons, which leads to the different relative momenta and
to some difference in spin-orbital interactions.

   \begin{figure}[ht!]
    \includegraphics[width=0.39\columnwidth]{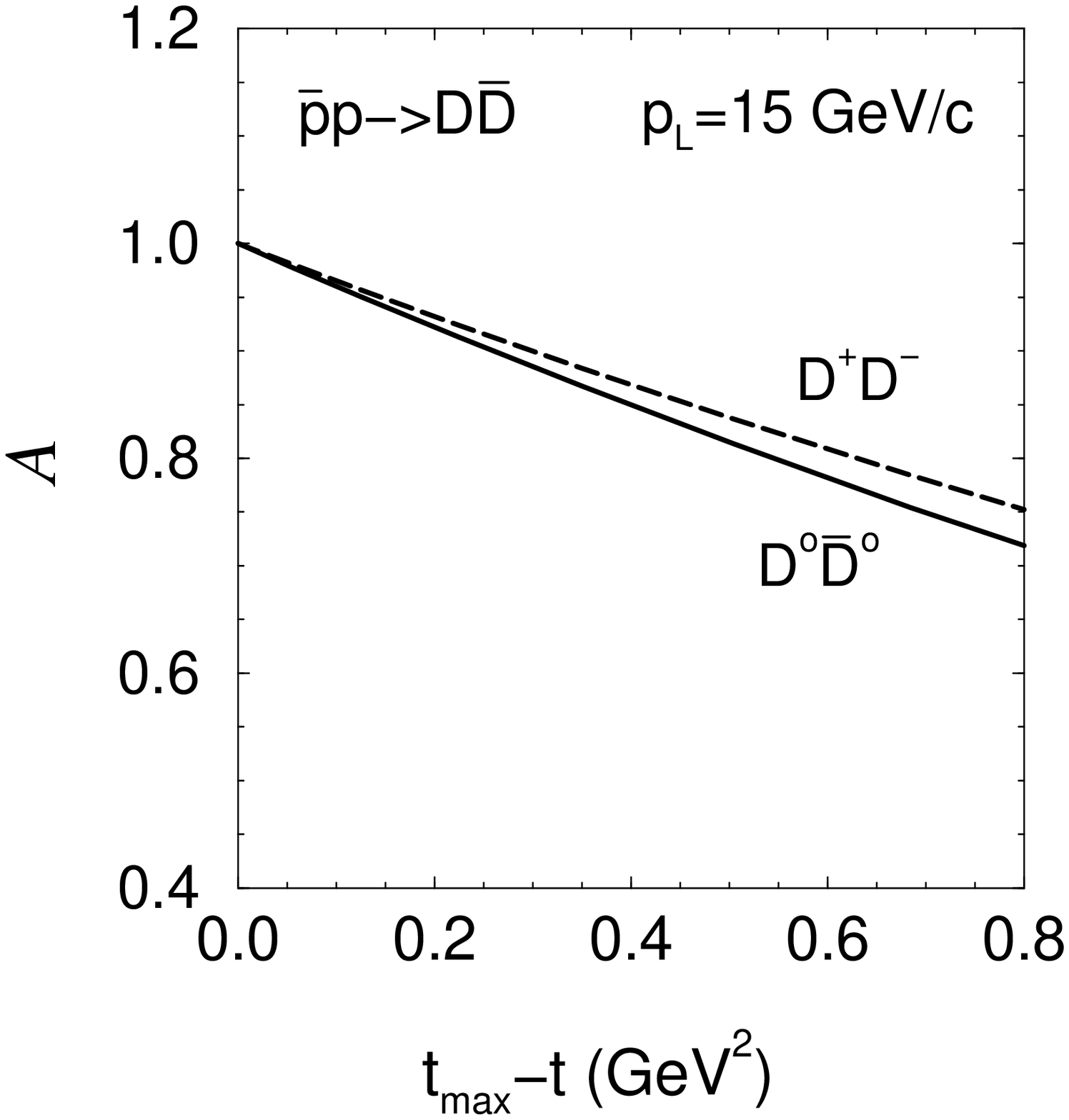}
    \qquad\qquad
    \includegraphics[width=0.35\columnwidth]{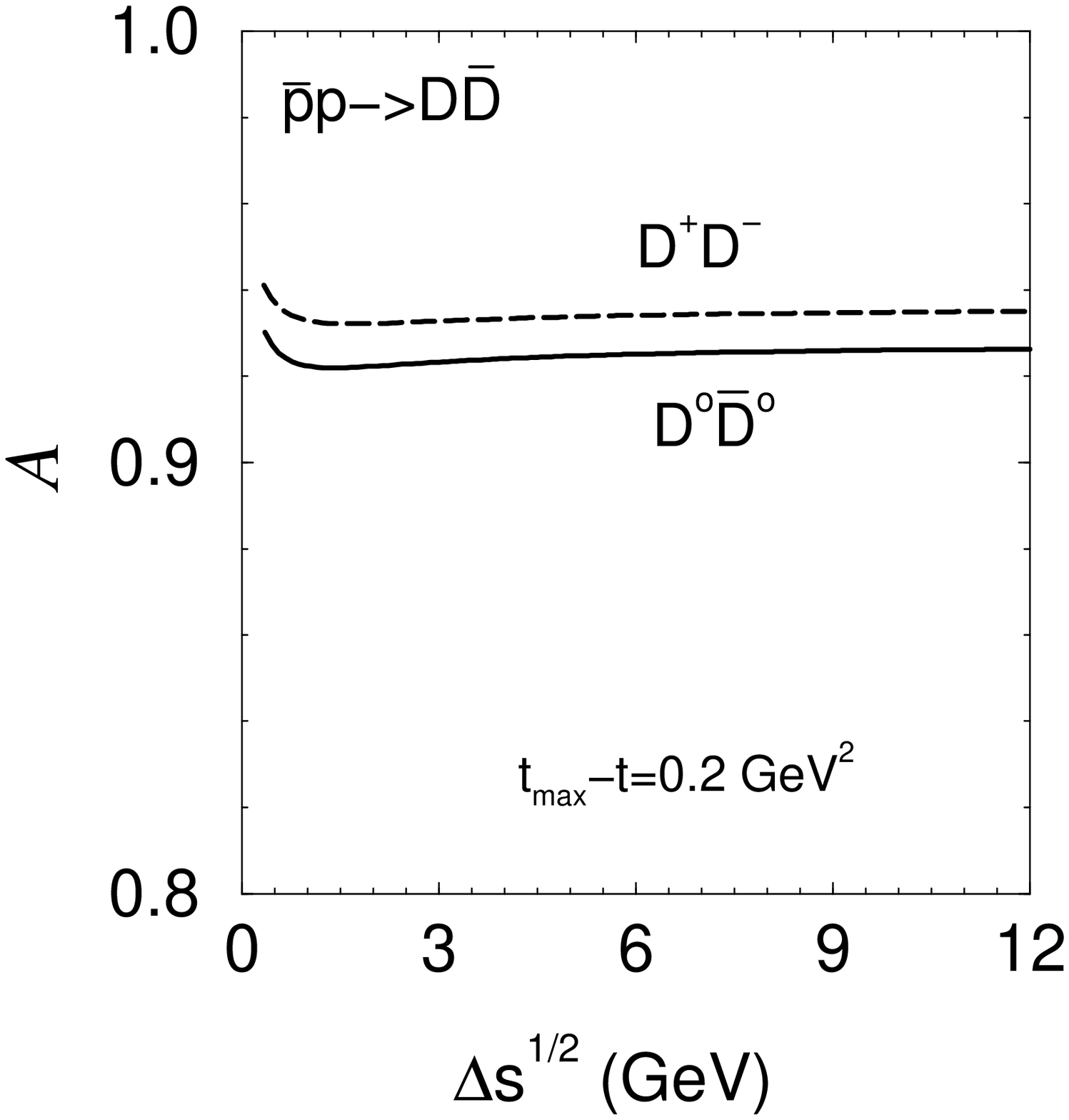}
   \caption{\small{
   The longitudinal asymmetry  for
   $\bar p p\to D^0\bar D^0$ (solid curves)
   and  $\bar p p\to D^-D^+$,(dashed curves).
  (Left panel: The asymmetry
   as a function of momentum transfer
   $t_{\rm max}-t$ at $p_L=15$~GeV.
   Right panel: The asymmetry  as a function of the energy
    excess $\Delta s^{1/2}$ at $t_{\rm max}-t=0.2$~GeV$^2$
\label{Fig:14}}}
  \end{figure}

The longitudinal asymmetries for the $\bar pp\to D\bar D$
reactions are presented in Fig.~\ref{Fig:14}. The left panel
demonstrates the $t$ dependence at the initial momentum
$p_L=15$~GeV/c. The right panel shows the dependence on the excess energy
at $t_{\rm max}-t=0.2$~GeV$^2$. One can see that the
results of the numerical calculation agrees with our above
qualitative consideration. The difference in the asymmetries for
$\bar KK$ and $D\bar D$ final states is mainly due to the
difference of masses of kaons and $D$ mesons.

\section{Reaction
$\boldmath \bar pp\to \bar MM^*$ }

The reactions $\bar pp\to \bar KK^*$ and  $\bar pp\to D\bar D^*$
are similar to the reactions with $\bar KK$ and $D\bar D$ final
states, and the corresponding amplitudes are described by the
diagrams depicted in Fig.~\ref{Fig:9}, where one of
the outgoing pseudoscalar mesons $M$ is replaced by the vector one, $M^*$,
i.e.\ $K\to K^*$, $D\to D^*$ etc.
Thus, the invariant amplitude for $\bar pp\to \bar KK$ reads
  \begin{eqnarray}
 T^{\bar pp\to\bar K^-K^{*+}}_{\lambda_f;m_i,n_i}
 =C'(t){\cal M}^{\bar pp\to\bar KK }_{\lambda_f;m_i,n_i}(s,t)
\frac{g_{K^*N\Lambda}g_{KN\Lambda} }{s_0}\,
\Gamma(\frac{1}{2}-\alpha_{ds}(t))\,
\left(-\frac{s}{s_{\bar pp:\bar KK^*}}
\right)^{\alpha_{ds}(t)-\frac{1}{2}},
 \label{E37}
 \end{eqnarray}
where $\lambda_f$ is the polarization of the outgoing $K^{*}$ and
the other notations have been introduced already in Sects.~II and
III. The baryon exchange trajectories are the same as in the previous
Sect.~III, $s_{\bar pp:\bar KK^*}=s_{\bar pp:\bar KK}$, and
$C'(t)$ is defined in Eq.~(\ref{E35})

The spin dependent amplitude ${\cal M}$ has the following form
\begin{eqnarray}
{\cal M}^{\bar pp\to\bar KK}_{\lambda_f;m_i n_i}(s,t) = {\cal
N}(s,t)\,\Gamma^{\mu}_{\lambda_f;m_in_i}
 \label{E38}
 \end{eqnarray}
with
\begin{eqnarray}
\Gamma^{\mu}_{\lambda_f;m_in_i}
 =\bar v_{n_i}\,\left[
\gamma_5\,(\fs p_Y - M_Y )\,
(\gamma^\mu+ \frac{\kappa_{NYK^*}}
{ 2(M_N+M_Y) }(\gamma^\mu\fs p_{K^*}
- \fs p_{K^* }\gamma^\mu) )\right]
u_{m_i}\epsilon^{\mu*}_{\lambda_i}~,
\label{E39}
\end{eqnarray}
where $\epsilon^{\mu}_{\lambda_i}$ is the polarization vector of
the $K^*$ meson and
\begin{eqnarray}
{\cal N}(s,t)= \frac{F_{\infty}(s)}{F(s,t)} \nonumber
 \end{eqnarray}
with
\begin{eqnarray}
 F^2(s,t)&=& {\rm Tr}
 \left[\Gamma_{\mu}{\Gamma_{\nu}}^\dagger \right] \, (-g^{\mu\nu} +
 \frac{p_{V}^\mu p_{V}^\nu}{M_V^2})~,\nonumber\\
 F^2_{\infty}&=& \frac{2sM_Y^2}{M_V^2} \left(
M_N^2 +M_V^2 + 6M_NM_V^2 z +M_V^2(2M_N^2+M_V^2)z^2\right)~,
\end{eqnarray}
where $M_V=M_{K^*}$, $p_V=p_{K^*}$ and $z=\kappa_{VNY}/(M_N+M_Y)$.

In the reaction $\bar pp\to K^-K^{*+}$ the total amplitude
is the coherent sum of $\Lambda$ and $\Sigma^0$ exchange channels.
In the case of $\bar K^0K^{*0}$, the amplitude is defined by the
$\Sigma^+$ exchange trajectory.

The amplitude for the $\bar K^*K$ reaction has a similar form:
\begin{eqnarray}
\Gamma^{\mu}_{\lambda_f;m_in_i}
 =\bar v_{n_i}\,\left[
(\gamma^\mu+ \frac{\kappa_{NYK^*}}
{ 2(M_N+M_Y) }(\gamma^\mu\fs p_{K^*}
- \fs p_{K^* }\gamma^\mu) )\,
(\fs p_Y - M_Y )\,
\gamma_5
\right]
u_{m_i}\epsilon^{\mu*}_{\lambda_i}~.
\label{E40}
\end{eqnarray}

The generalization for $D\bar D^*$ and $D^*\bar D$ final states
may be done in a straightforward manner, similarly to the previous Sections.\\

{\it 1.~~~Differential cross sections}\\

\begin{figure}[ht]
    \includegraphics[width=0.35\columnwidth]{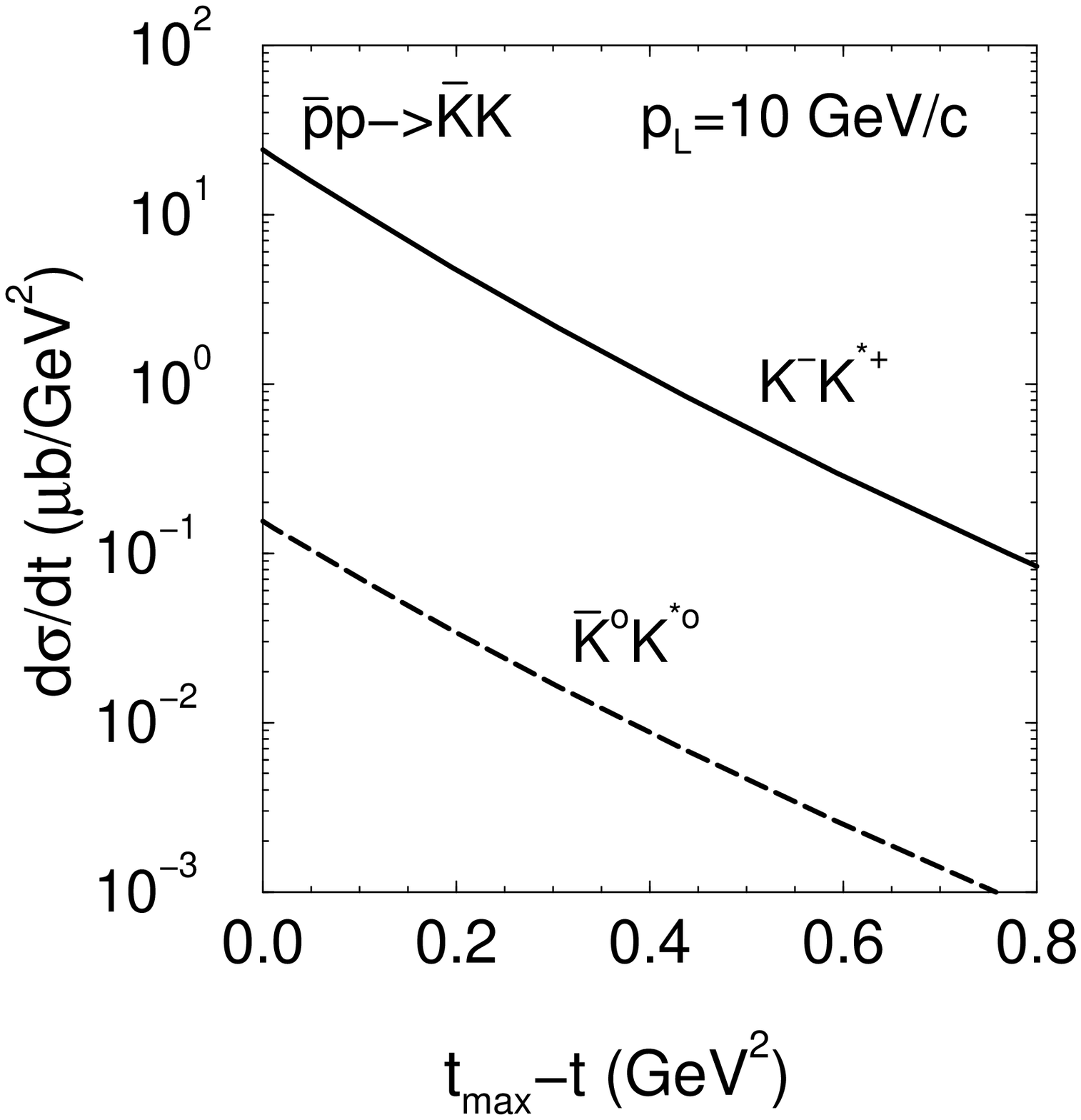}
    \qquad\qquad
    \includegraphics[width=0.35\columnwidth]{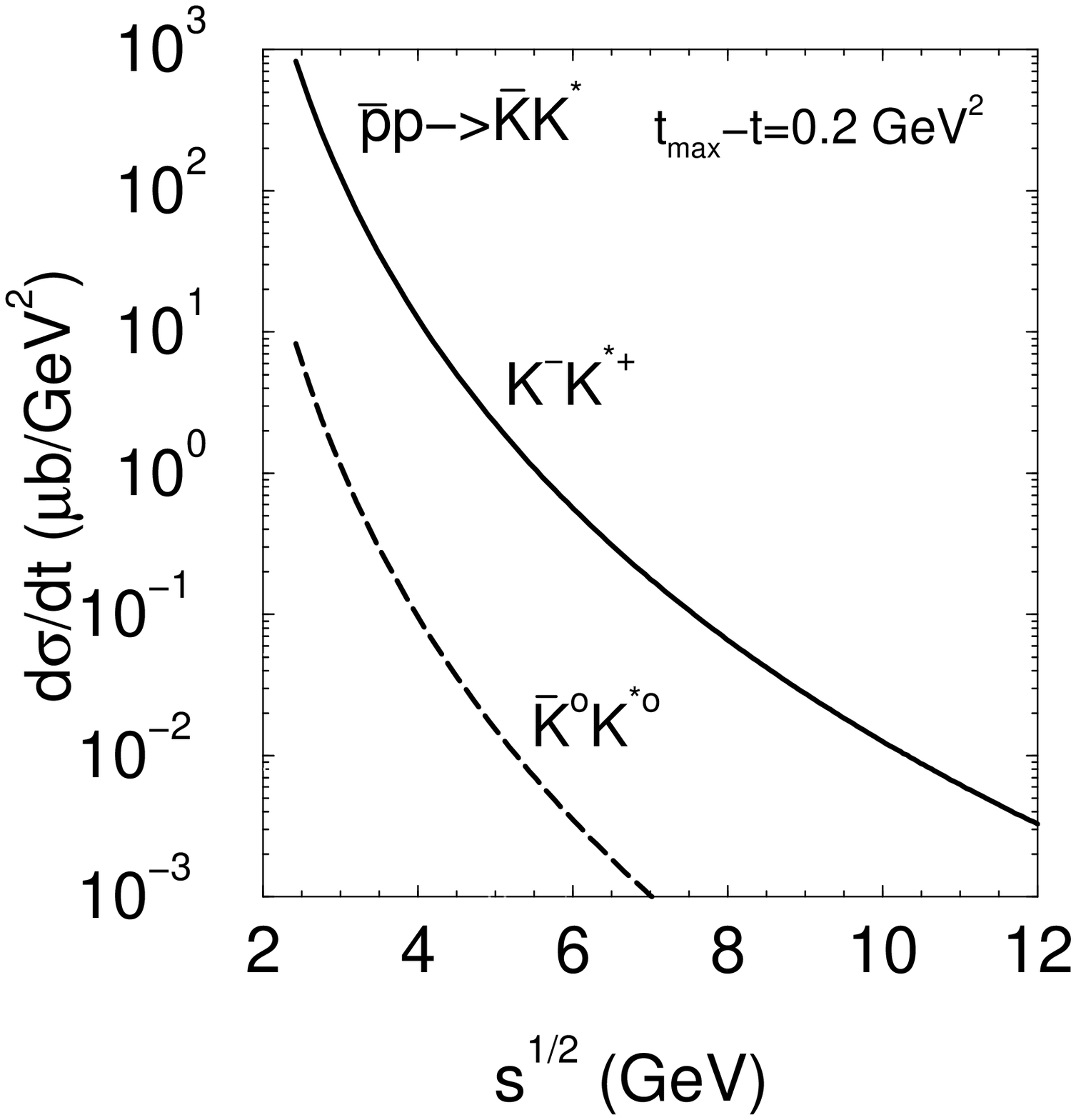}
   \caption{\small{Left panel:
   The differential cross sections
   of the $\bar p p\to K^-K^{*+}$
   (solid curve) and $\bar pp\to\bar K^0K^)$ (dashed curve)
   reactions as a function
  of $t_{\rm max}-t$ at $p_L=10$~GeV/c.
 Right panel: The differential cross sections
  as a function of the energy
  $\sqrt{s}$ at $t_{\rm max}-t=0.2$~GeV$^2$.
\label{Fig:15}}}
  \end{figure}

 The differential cross sections of the reactions
 $\bar pp\to\bar KK^*$ and $\bar pp \to \bar K^*K)$
 are exhibited in Fig.~\ref{Fig:15}.
 The left panel shows our prediction for
 the differential cross sections
 of the reactions $\bar p p\to  K^-K^{*+}$,
 and $\bar p p\to \bar K^0K^{*0}$
 as a function of $t_{\rm max}-t$ at initial momentum
 $p_L=10$~GeV/c.
 The dependence of the differential cross sections
 on energy $\sqrt{s}$ at fixed $t_{\rm max}-t=0.2$~GeV$^2$
 is presented in the right panel.
 At large energies the cross sections behaves similarly
 to the cross sections of the $\bar pp\to\bar KK$ reactions.

\begin{figure}[h!]
    \includegraphics[width=0.35\columnwidth]{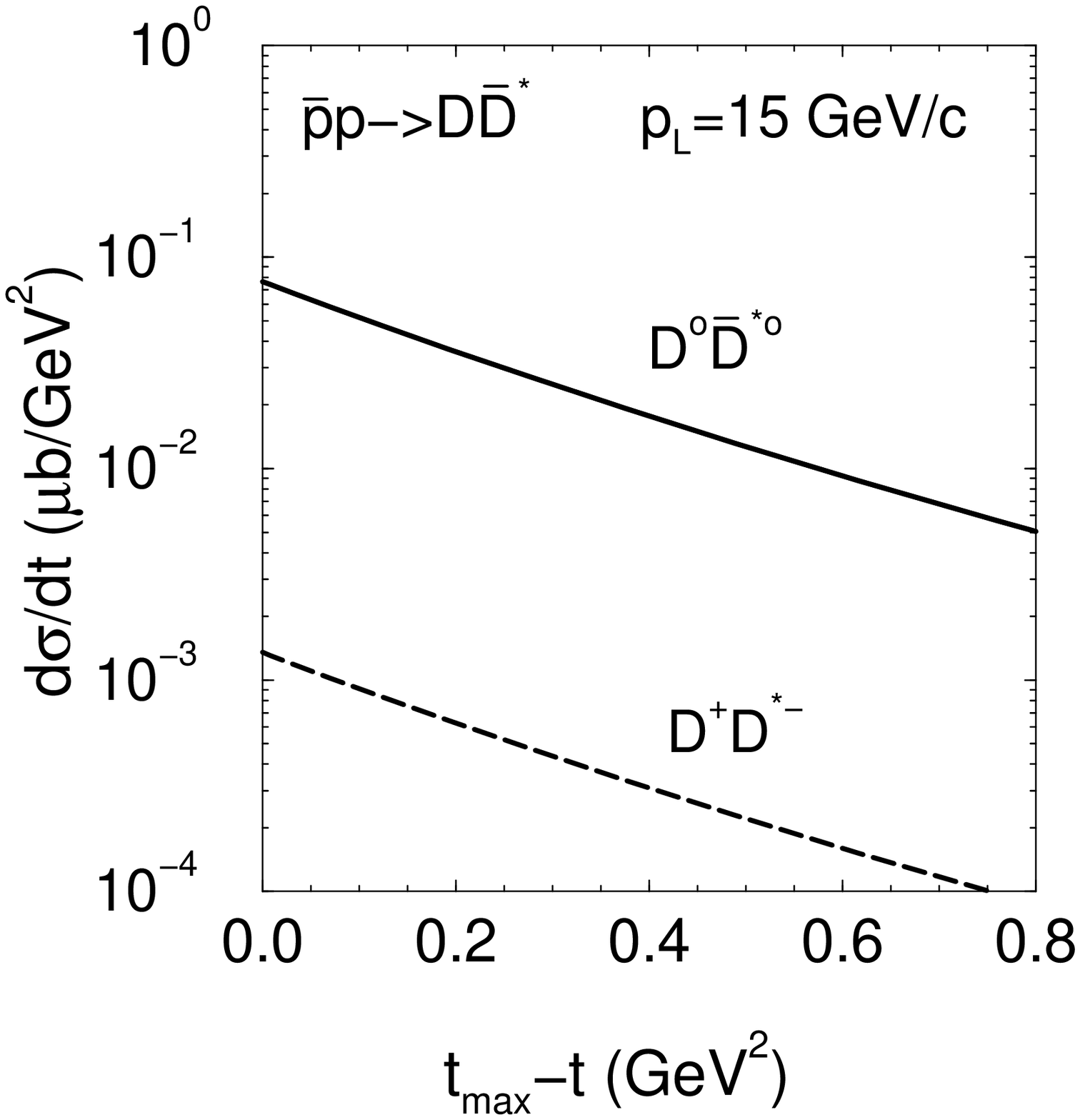}
    \qquad\qquad
    \includegraphics[width=0.35\columnwidth]{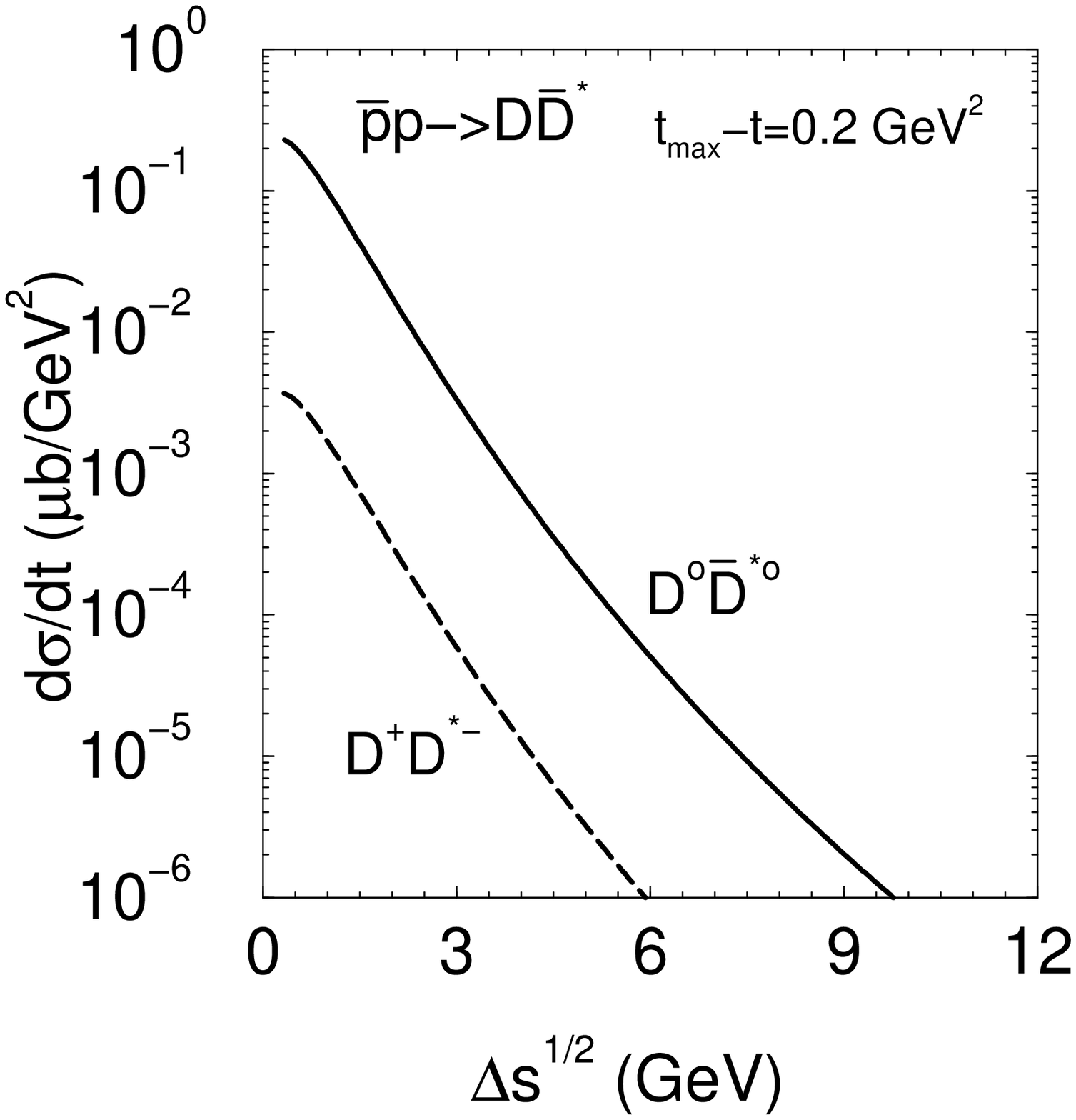}
   \caption{\small{Left panel:
  The differential cross sections of the
  $\bar p p\to \bar D^0D^{*}$
   (solid curve) and $\bar pp\to D^-D^{*+}$ (dashed curve)
  reactions as a function of $t_{\rm max-t}$ at $p_L=15$~GeV/c.
 Right panel: Differential cross section as
 a function of the energy excess
 $\Delta s^{1/2}$ at $t_{\rm max}-t=0.2$~GeV$^2$.
\label{Fig:16}}}
  \end{figure}

Our prediction for the differential cross sections
of the $D\bar D^*$ pair production
is presented in Fig.~\ref{Fig:16}.
The left panel illustrates the dependence on $t_{\rm max} - t$ at
fixed $p_L=15$~GeV/c. The right panel exhibits the dependence on
the energy excess $\Delta s^{1/2}$ at fixed $ t_{\rm
max}-t=0.2$~GeV$^2$. The ratio of the cross sections with
$D^-D^{*+}$ and   $\bar D^0D^{^0}$ final states is defined by the
coupling constants in $KN\Sigma,\,{K^*N\Sigma}$ and
${KN\Lambda},\,{K^*N\Lambda}$ interactions and is close to 0.03.
The cross sections decrease with energy similarly to the $\bar
pp\to D\bar D$ reactions, and therefore, the region with small
excess energy is more suitable for studying these reactions.\\

{\it 2.~~~Longitudinal asymmetry}\\

Let us consider first the case of the forward production angle (or
$t=t_{\rm max}$), in $\bar pp\to\bar KK^*$ ($D\bar D^*$) reactions
with pure vector coupling in ${V^*N\Sigma}$ ($V=K^*,\,D^*$). In
case of large energies, where $\bf{p}_V\simeq \bf{p}_p $, the
amplitude  has the form
\begin{eqnarray}
T_{\lambda_i;m_i,n_i}\sim R(s) \left(
A\,\delta_{m_in_i}
+B\,\,\delta_{-m_in_i}\right)\delta_{\lambda_i\lambda_V}~,
\label{E41}
 \end{eqnarray}
where $\lambda_V$ is the polarization of the outgoing
vector meson, $\lambda_i=m_i+n_i$, and
\begin{eqnarray}
A\simeq \sqrt{2},\qquad
B\simeq \frac{M_N}{M_V}~,
\label{E42}
\end{eqnarray}
where $M_V$ is the mass of the vector meson. This results in
 \begin{eqnarray}
 {\cal A}= \frac{M_N^2-2M_V^2}{M_N^2+2M_V^2}~.
\label{E43}
 \end{eqnarray}
Thus, for the $\bar KK^*$ final state, where $M_V\sim M_N$ the
asymmetry has values ${\cal A}\simeq-0.3$ and increases at finite
production angles because of the spin-orbital interaction. In case
of the $\bar pp\to D\bar D^*$) reaction, the asymmetry is much
smaller: ${\cal A}\simeq-0.8$ and it also increases with the
production angle. The finite tensor coupling changes these
predictions, especially in the charm sector with large $M_V$.

    \begin{figure}[ht!]
    \includegraphics[width=0.35\columnwidth]{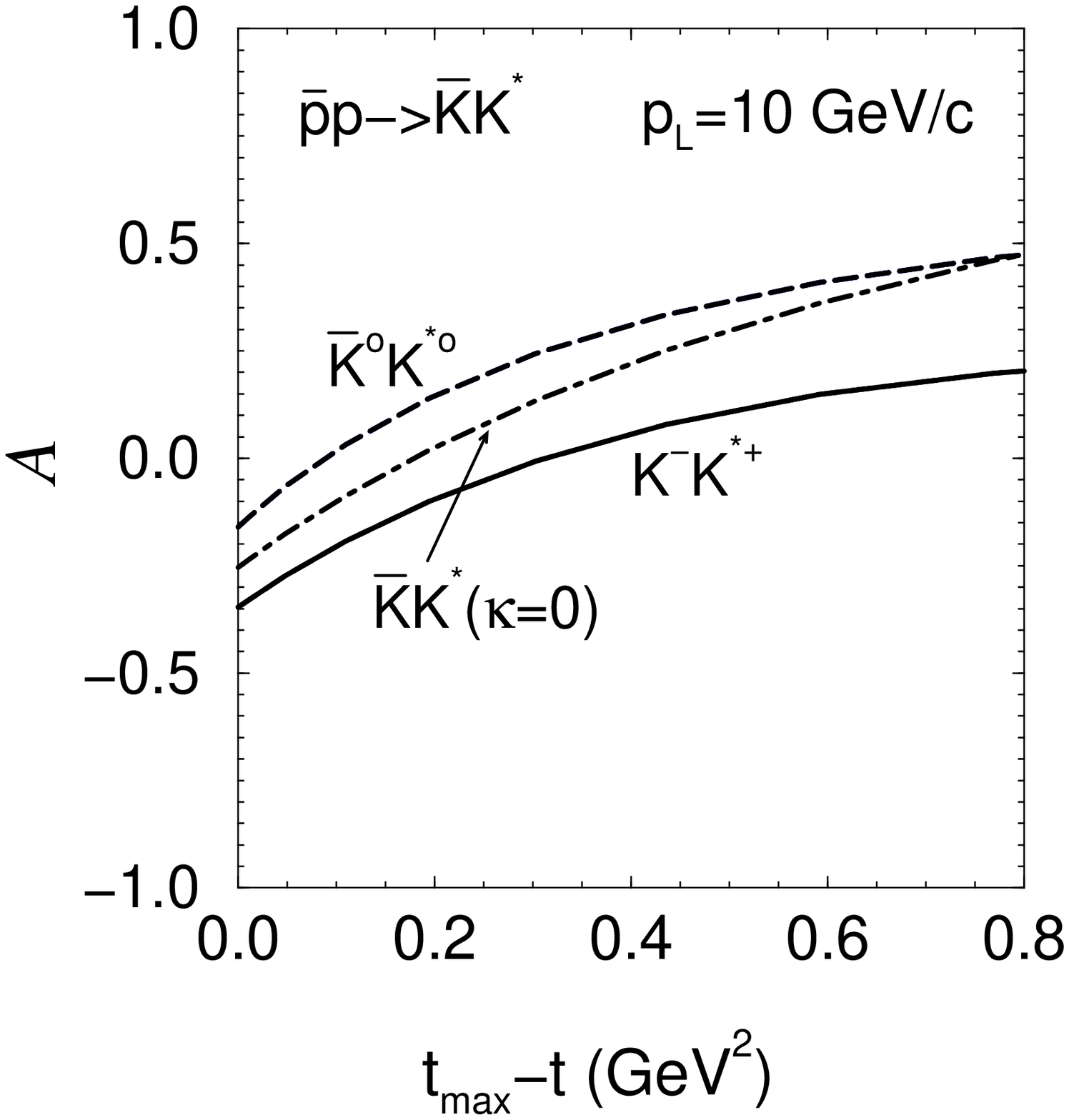}
    \qquad\qquad
    \includegraphics[width=0.35\columnwidth]{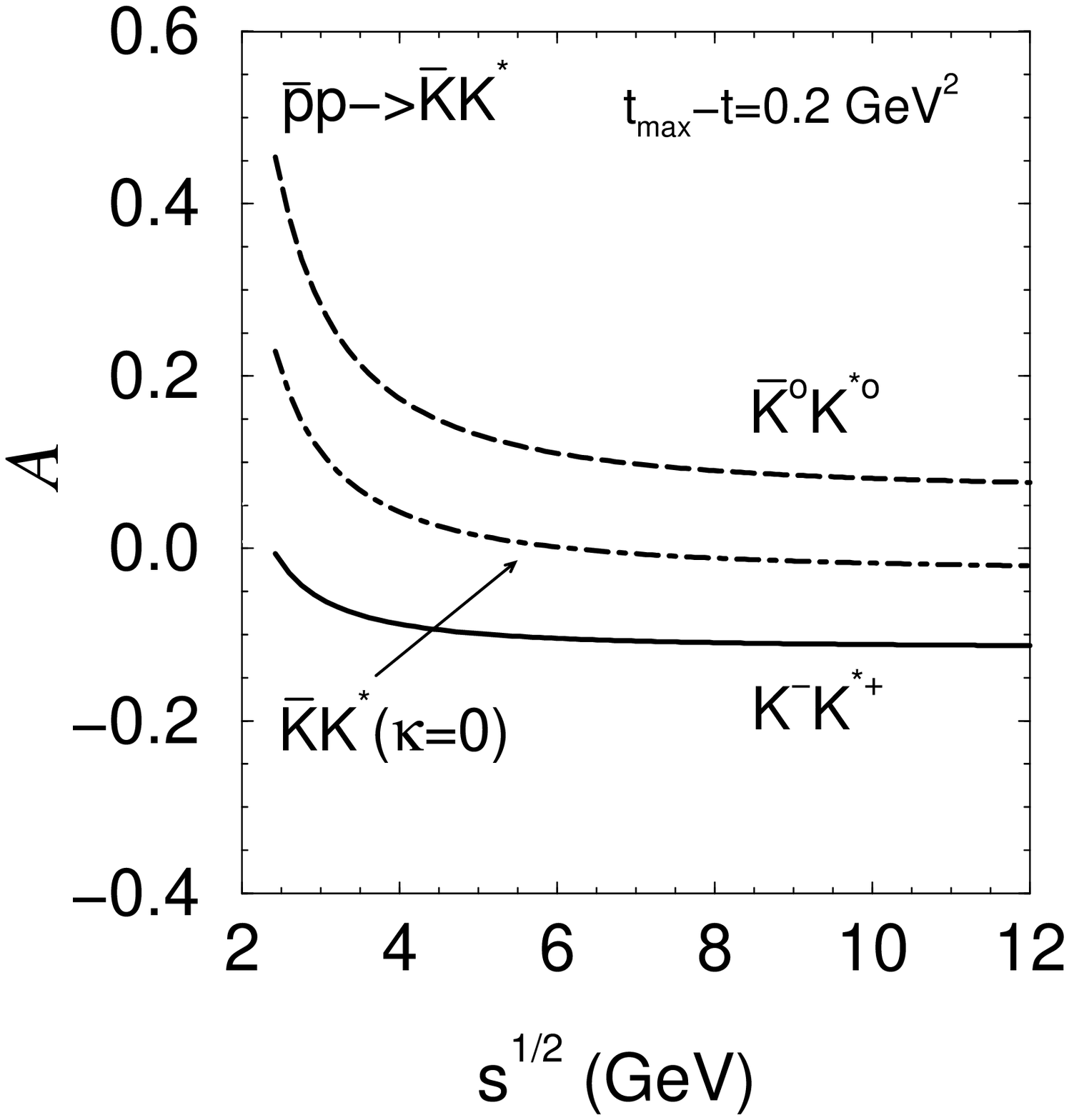}
   \caption{\small{
   The longitudinal asymmetry for the reactions
   $\bar p p\to K^-K^{*+}$ (solid curves)
   and  $\bar p p\to \bar K^0K^{*0}$, (dashed curves).
   The dot-dashed curves correspond to the
   calculation with zero tensor coupling constant.
  Left panel: The asymmetry
   as a function of momentum transfer
   $t_{\rm max}-t$ at $p_L=10$~GeV.
   Right panel: The asymmetry  as a function of the energy
    $\sqrt{s}$ at $t_{\rm max}-t=0.2$~GeV$^2$.
\label{Fig:17}}}
  \end{figure}

In Fig.~\ref{Fig:17} we show our prediction for the $\bar
pp\to\bar KK^*$ reactions. The left panel exhibits the $t$ dependence at
$p_L = 10$~GeV/c, whereas the right panel depicts the
$\sqrt{s}$ dependence at $t_{\rm max}-t=0.2$~GeV$^2$. One can see
an increase of the asymmetry with $-t$ and its almost constant value
at large $\sqrt{s}$ and fixed $t_{\rm max}-t$. The difference in
${\cal A}$ for $\bar K^0K^{*0}$ and $K^+K^{*-}$ final states is
mainly due to the difference in  tensor couplings in
${K^*N\Sigma}$ and ${K^*N\Lambda}$ interactions. For completeness,
we also show the result for a calculation without tensor couplings. At
zero production angle the asymmetry is close to our above qualitative
estimate.

   \begin{figure}[ht!]
    \includegraphics[width=0.39\columnwidth]{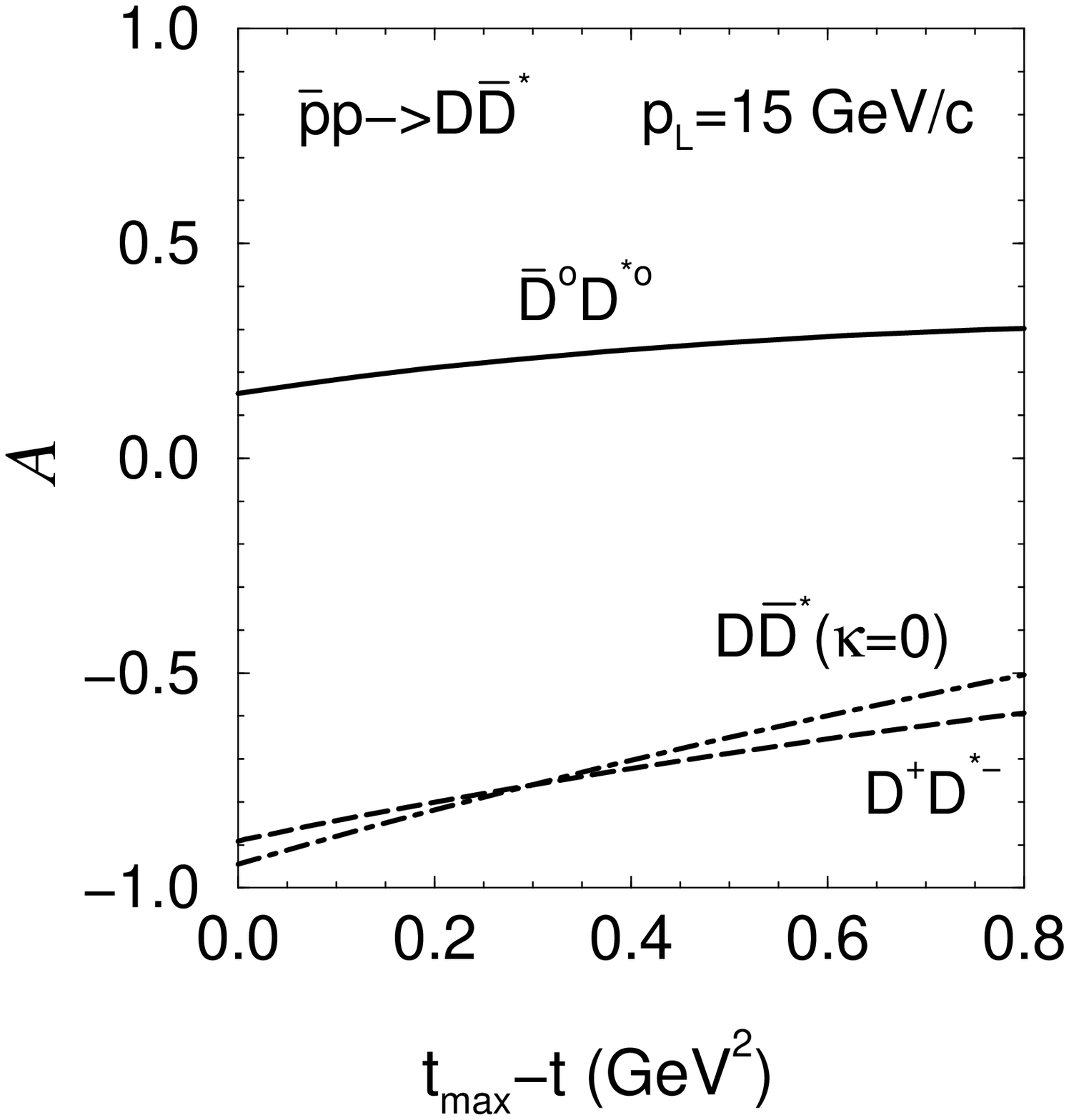}
    \qquad\qquad
    \includegraphics[width=0.35\columnwidth]{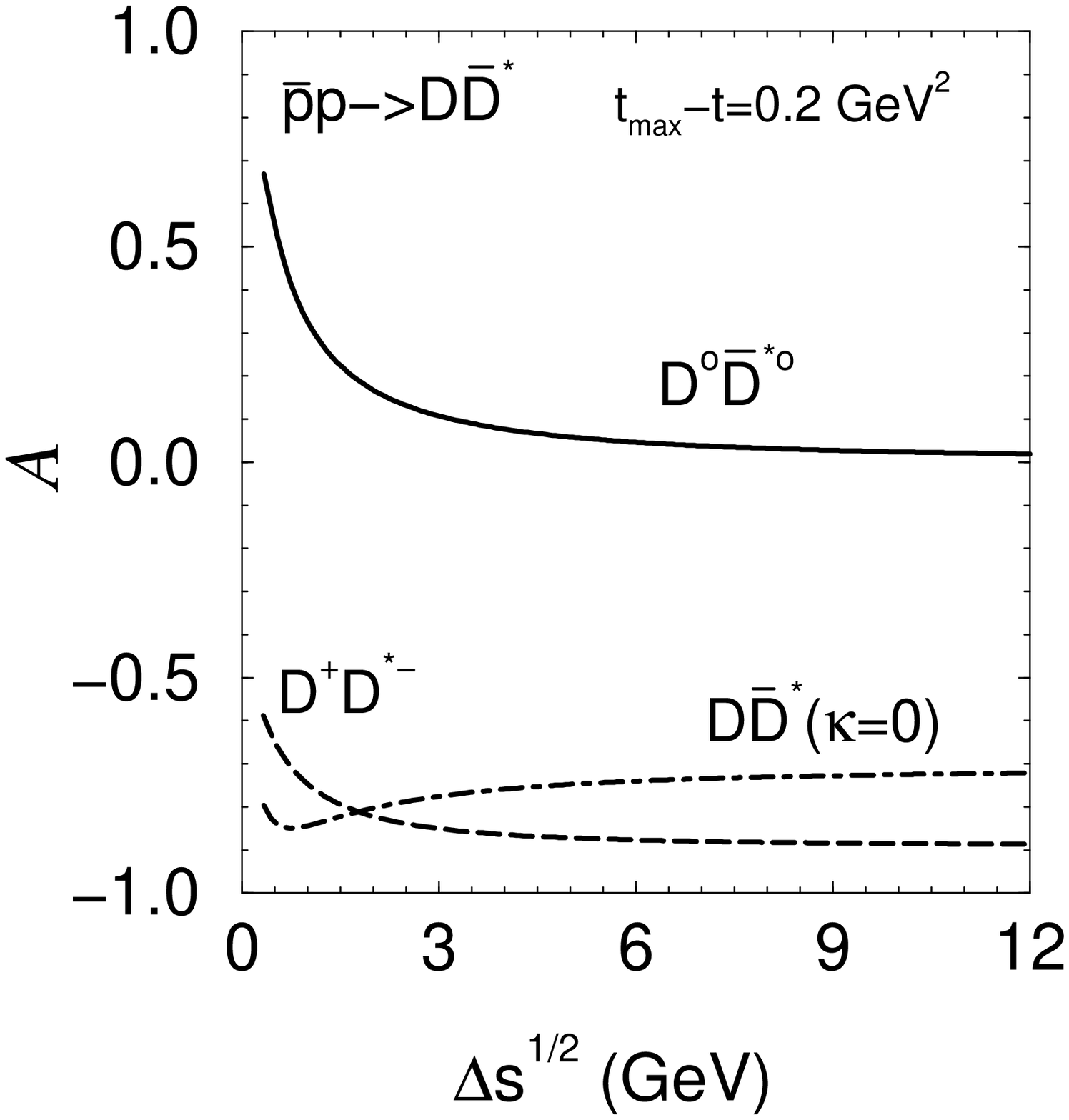}
   \caption{\small{
   The longitudinal asymmetry  for reactions
   the $\bar p p\to D^0\bar D^{*0}$ (solid curves)
   and  $\bar p p\to D^-D^+$ (dashed curves).
  The dot-dashed curves correspond to the
   calculation with zero tensor coupling constant.
  Left panel: The asymmetry
   as a function of momentum transfer
   $t_{\rm max}-t$ at $p_L=15$~GeV.
   Right panel: The asymmetry  as a function of the excess energy
    $\Delta\sqrt{s}$ at $t_{\rm max}-t=0.2$~GeV$^2$
\label{Fig:18}}}
\end{figure}

The longitudinal asymmetries for $\bar pp\to D\bar D^*$
are presented in Fig.~\ref{Fig:18}. The left panel
displays the $t$ dependence at initial momentum $p_L=15$~GeV/c.
The right panel shows the dependence on the energy excess at
$t_{\rm max}-t=0.2$~GeV$^2$. We also show results for a calculation
with pure vector couplings. At zero production angle the asymmetry
(left panel) coincides with our qualitative estimate (cf.\
Eq.~(\ref{E43})). The effect of the tensor interaction is rather large.
One can see a big difference between $D^0D^{*0}$ and $D^-D^{*+}$
final states because of the difference in the corresponding tensor
couplings. Numerically, result for the $D^-D^{*+}$ final state is
close to a calculation with zero tensor coupling.

\section{Summary}

In summary, we have analyzed the open charm production in the
exclusive binary reactions $\bar pp\to \bar Y_cY_c$, $\bar pp\to
D\bar D$ and $\bar pp\to D\bar D^*$ at small momentum transfer.
Our consideration is based on a modified Regge type model,
motivated by quark-gluon string dynamics. The most important
parameters of the model are the effective charmed meson and baryon
exchange trajectories and the energy scale parameters. They are
found from a consistent approach based on the topological
decomposition and  factorization of the corresponding planar quark
diagrams. The coupling constants are taken to be the same as in
corresponding strangeness production reactions, assuming SU(4)
symmetry. Unknown residual functions are found from the comparison
of the calculation of $\bar pp\to
\bar\Lambda\Lambda\,(\bar\Lambda\Sigma)$ and $\bar pp\to \bar KK$
reactions with available (although old) experimental data. As a
result, we obtained the absolute value of the corresponding cross
sections in the energy range of future FAIR experiments.

For the first time we made predictions for the longitudinal asymmetry, 
which is quite different in different processes with
non-trivial $t$ and $s$ dependencies. In each case we presented an
analytical estimate for the forward production 
with the aim to understand the physics of the asymmetry.

Our calculations of cross sections and longitudinal
asymmetries in
the exclusive reactions $\bar pp\to \bar YY$, $\bar pp\to \bar KK$ and
$\bar pp\to \bar KK^*$  have also an independent interest for
forthcoming experiments at FAIR as a first prediction of the open strangeness
production in peripheral reactions in this energy region.
Our consideration may serve as a first step towards more
involved reaction mechanisms and extensions to $pp$ collisions.

 \acknowledgments

 One of the authors (A.I.T.) appreciates the warm
 hospitality in Forschungszentrum Dresden-Rossendorf.
 The work was supported by BMBF grant 06DR136 and GSI-FE.

 \end{document}